\documentclass[aps,prb,reprint,amsmath,amssymb,superscriptaddress]{revtex4-2}
\usepackage{amsmath} %
\usepackage{bm}  %
\usepackage{graphicx}
\usepackage{physics}
\usepackage{color}
\usepackage{booktabs}
\usepackage[colorlinks=true, citecolor=blue, linkcolor=blue, urlcolor=blue]{hyperref}
\usepackage{soul}
\graphicspath{ {./}{./figures/} }

\newcommand{\kk}{\bm{k}}
\renewcommand{\qq}{\bm{q}}
\newcommand{\pp}{\bm{p}}

\renewcommand{\vec}[1]{\mathbf{#1}}

\begin{document}

\title{Many-body theory of phonon-induced spin relaxation and decoherence}%
\author{Jinsoo Park}
\affiliation{Department of Applied Physics and Materials Science, California Institute of Technology, Pasadena, CA 91125, USA.}
\author{Yao Luo}
\affiliation{Department of Applied Physics and Materials Science, California Institute of Technology, Pasadena, CA 91125, USA.}
\author{Jin-Jian Zhou}
\affiliation{School of Physics, Beijing Institute of Technology, Beijing 100081, China.}
\author{Marco Bernardi}
\email[Corresponding author: ]{bmarco@caltech.edu}
\affiliation{Department of Applied Physics and Materials Science, California Institute of Technology, Pasadena, CA 91125, USA.}

\begin{abstract}
First-principles calculations enable accurate predictions of electronic interactions and dynamics. 
However, computing the electron spin dynamics remains challenging. 
The spin-orbit interaction causes various dynamical phenomena that couple with phonons, such as spin precession and spin-flip $e$-ph scattering, which are difficult to describe with current first-principles calculations. %
In this work, we show a rigorous framework to study phonon-induced spin relaxation and decoherence, by computing the spin-spin correlation function and its vertex corrections due to $e$-ph interactions. 
We apply this approach to a model system and develop corresponding first-principles calculations of spin relaxation in GaAs.
Our vertex-correction formalism is shown to capture the Elliott-Yafet, Dyakonov-Perel, and strong-precession mechanisms $-$ three independent spin decoherence regimes with distinct physical origins $–$ thereby unifying their theoretical treatment and calculation.
Our method is general and enables quantitative studies of spin relaxation, decoherence, and transport in a wide range of materials and devices. 
\end{abstract}
\maketitle
\section{Introduction}
\vspace{-10pt}
Linear response theory provides a microscopic understanding of the response of a system to external perturbations and computes the associated correlation functions ~\cite{mahanManyParticle2000,rohlfingElectronhole2000,sekineQuantum2017,kimVertex2019,zhouPredicting2019,desaiMagnetotransport2021}. First-principles calculations of electronic interactions~\cite{baroniPhonons2001,giustinoElectron2007,agapitoAbInitio2018,jhalaniPiezoelectric2020,bruninElectronphonon2020,luAbInitio2020,LuFirst2021,zhouInitio2021} complement this formalism, enabling precise predictions of materials properties and transport coefficients without resorting to empirical models or fitting parameters.
In this context, electron-phonon ($e$-ph) interactions are particularly important as they govern a wide range of phenomena such as charge transport~\cite{bernardiFirstprinciples2016}, superconductivity~\cite{pickettElectronic1989}, spin transport~\cite{balasubramanianNanoscale2008,jansenSilicon2012,zuticSpintronics2004} and spin decoherence~\cite{veldhorstTwoqubit2015,noiriFast2022,petitUniversal2020}.
\\
\indent
The Boltzmann transport equation (BTE) is widely used to study the response to an external electric field~\cite{Ziman,Lundstrom}.  
The field drives the electronic populations $f_{n\kk}$, for states with band $n$ and crystal momentum $\kk$, away from the equilibrium Fermi-Dirac distribution $f^0_{n\kk}$, while the $e$-ph interactions dissipate electron energy and act to restore equilibrium, resulting in a steady-state current $e(f_{n\kk}-f^0_{n\kk})v_{n\kk}$, where $e$ is the electron charge and $v_{n\kk}$ is the band velocity~\cite{Ziman}. 
In the many-body formalism, the BTE at low electric field is formally equivalent to the ladder vertex-correction to the dc conductivity~\cite{kimVertex2019}. 
In that framework, one determines the current-current correlation function, with vertex corrections from the $e$-ph interactions obtained by summing over ladder diagrams, and computes the conductivity from the dissipative part of the susceptibility.
A key factor making this approach equivalent to the BTE is that the electron velocity is band-diagonal in the Bloch basis, $\bra{m\kk}\hat{v}\ket{n\kk}=\delta_{nm}\partial_{\kk} E_{n\kk}/\hbar$~\cite{kimVertex2019}. 
\\
\indent
However, studying the response to an external field of an arbitrary operator that couples with phonons is more difficult. 
The matrix representation of an operator $\hat{A}$ is in general nondiagonal in the Bloch basis, $\hat{A}_{nm\kk}\!=\!\bra{m\kk}\hat{A}\ket{n\kk}$, and can mix states in \mbox{different bands.} 
The BTE cannot be applied in this case because due to its population-based formalism it neglects such off-diagonal (inter-band) components. A framework treating the response of non-diagonal operators coupled with $e$-ph interactions is still missing. %
\\
\indent
An important example is spin relaxation and decoherence, where spin-orbit coupling (SOC) makes the spin operators non-diagonal in the band index, and phonons can change the electron spin through $e$-ph interactions~\cite{zuticSpintronics2004}.
Theories of spin decoherence focus on two distinct models $-$ the Elliott-Yafet (EY) mechanism~\cite{elliottTheory1954,yafetFactors1963}, where $e$-ph collisions rotate the spin direction, and the Dyakonov-Perel (DP) mechanism~\cite{dyakonov1972spin}, where spin precession in the SOC field induces a motional narrowing of the spin.
The dominant mechanism depends on the system $-$ typically, EY dominates in centrosymmetric and DP in non-centrosymmetric materials. 
Spin relaxation exhibits opposite trends in these two mechanisms, with spin relaxation times  proportional to the $e$-ph relaxation times in EY, and inversely proportional in DP.
We have recently shown that EY spin relaxation can be computed from first-principles in the spin relaxation time approximation (sRTA)~\cite{parkSpinphonon2020}$-$the spin counterpart of the transport RTA for charge transport~\cite{mahanManyParticle2000,kimVertex2019} $-$ but spin precession and the DP mechanism are neglected in the sRTA.
\\
\indent %
Here we show a many-body approach to compute the susceptibility for an arbitrary non-diagonal operator coupled to $e$-ph interactions.
Our diagrammatic approach, based on the Kubo formula with vertex corrections to the susceptibility in an external injection field, calculates an effective phonon-dressed operator and its renormalized dynamics. 
We derive a Bethe-Salpeter equation (BSE) for the vertex corrections, and specializing to the spin operator, we use the vertex corrections to compute spin relaxation and precession. 
We show that the vertex corrections can capture spin decoherence due to both the EY and DP mechanisms and can also model the strong-precession regime, a third mechanism distinct from EY and DP. 
We find these three mechanisms in the exact solution of a two-level system, and also identify them in a real material, GaAs, using first-principles calculations. %
Combined with first-principles $e$-ph calculations, our method is poised to advance microscopic understanding of phonon-induced spin decoherence~\cite{parkPredicting2022}, with applications ranging from solid-state qubits to quantum materials with spin Hall effect, valley-dependent spin physics, and Rashba effect.  
\\
\indent
The paper is organized as follows: 
In Sec.~\ref{sec:Theory}, we derive the BSE for the phonon-dressed vertex, discuss its physical interpretation, %
and calculate the susceptibility in response to an injection field.
In Sec.~\ref{sec:spin}$-$\ref{sec:Results}, we apply this formalism to study spin dynamics in a model two-level system and in a real material, GaAs, discussing spin relaxation due to the EY, DP, and strong-precession mechanisms.\\
\section{Theory}\label{sec:Theory}
We derive a self-consistent BSE for the vertex correction to the susceptibility due to $e$-ph interactions, focusing on a general vector observable $\mathbf{\hat{A}}$.
We then present a physical interpretation of the vertex corrections and the renormalized dynamics of the operator. 
We employ atomic units and set $\hbar=1$.
\subsection{Interacting Green's function}
\vspace{-8pt}
We consider an unperturbed Hamiltonian $H_0$ diagonal in the Bloch basis, $\bra{n'\kk}H_0\ket{n\kk}=\varepsilon_{n\kk}\delta_{nn'}$.
The interacting imaginary-time Green's function $\mathcal{G}(i\omega_a)$ is written using the Dyson equation as~\cite{mahanManyParticle2000}
\begin{equation}
\mathcal{G}(i\omega_a)^{-1}=\mathcal{G}^{(0)}(i\omega_a)^{-1}-\Sigma(i\omega_a),
\end{equation}
where $\omega_a$ are fermionic Matsubara frequencies, 
$\mathcal{G}^{(0)}(i\omega_a)%
$ is the non-interacting Green's function, and 
$\Sigma(i\omega_a)$ is the lowest order (Fan-Migdal) $e$-ph self-energy~\cite{mahanManyParticle2000,bernardiFirstprinciples2016,cardonaRenormalization2001}, whose band- and $\kk$-dependent expression is
\begin{equation} \label{eq:sigma}
\begin{split}
\Sigma_{nn'\kk}&(i\omega_a)=-\frac{1}{\beta N_q V_{\text{uc}}}\sum_{mm'\qq\nu,iq_c} \left[g_{n'm'\nu}(\kk,\qq)\right]^* g_{nm\nu}(\kk,\qq) \\
&\times\mathcal{D}_{\nu\qq}(iq_c)\mathcal{G}_{mm'\kk+\qq}(i\omega_a+iq_c).
\end{split}
\end{equation}

Here, $\beta=1/{k_B T}$ at temperature $T$, $N_q$ is the number of $\qq$-points in the summation, $V_{\text{uc}}$ is the unit cell volume, $q_c$ is the bosonic Matsubara frequency of the phonon, and $\mathcal{D}_{\nu\qq}(iq_c)=2\omega_{\nu \qq}/((iq_c)^2-\omega_{\nu \qq}^2)$ is the non-interacting phonon Green's function for a phonon with mode index $\nu$, wave-vector $\qq$, and energy $\omega_{\nu\qq}$. 
The $e$-ph matrix elements $g_{nm\nu}(\kk,\qq)$ quantify the probability amplitude for an electron in a Bloch state $\ket{\psi_{n\kk}}$, with band index $n$ and crystal momentum $\kk$, to scatter into a final state $\ket{\psi_{m\kk+\qq}}$ 
by emitting or absorbing a phonon~\cite{bernardiFirstprinciples2016,zhouPerturbo2021},
\begin{equation} \label{eq:me}
g_{nm\nu}(\kk,\qq) \!=\! \bra{\psi_{m\kk+\qq}} 
\partial_{ \nu \qq} \hat{V}\! \ket{\psi_{n\kk}}\!, 
\end{equation}
where $\partial_{\nu \qq} \hat{V}$ is the perturbation to the potential acting on an electron due to a given phonon mode $(\nu,\qq)$.  
\begin{figure}[!t]
\includegraphics[width=0.95\columnwidth]{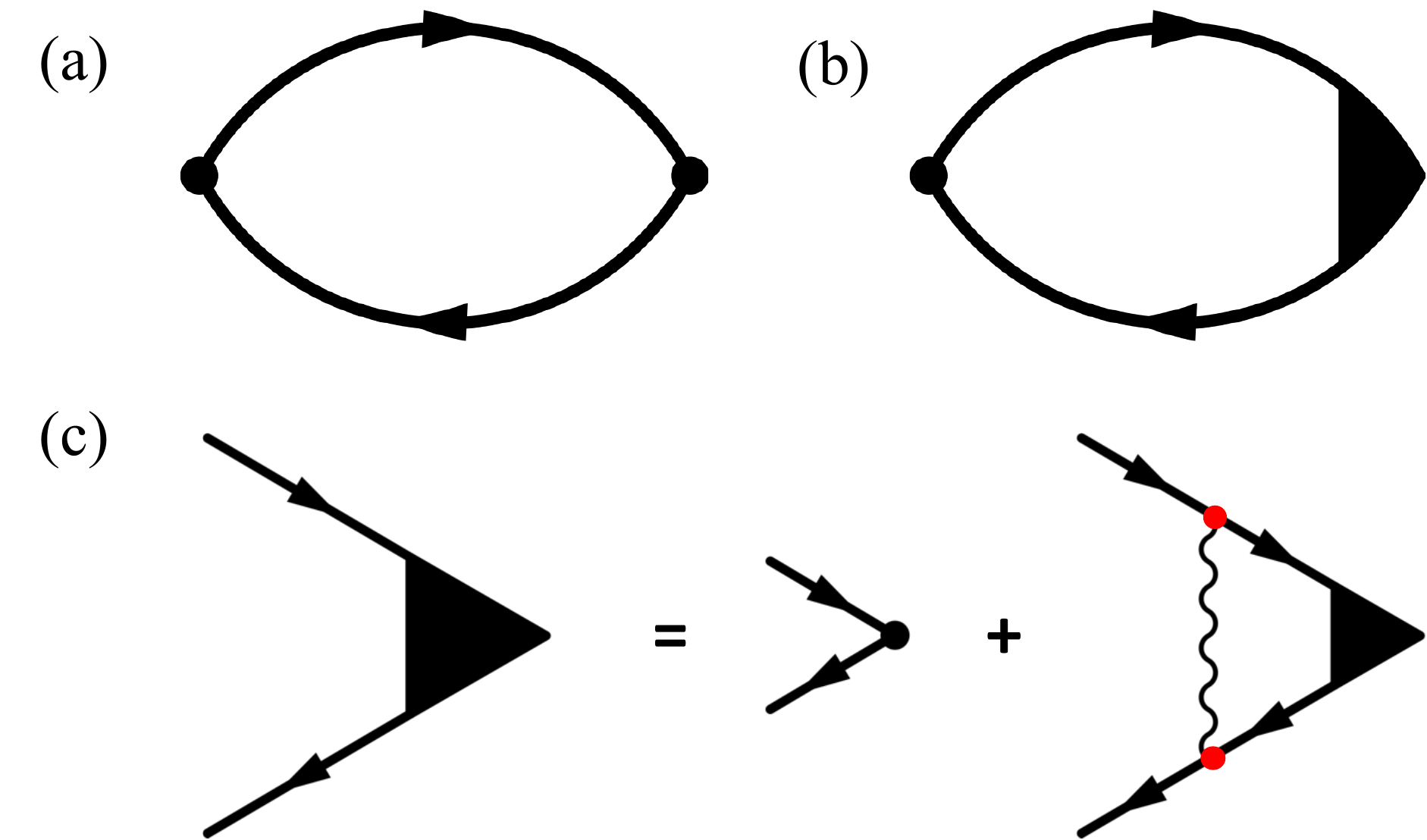}
\caption{(a) Bare bubble diagram without the vertex correction.
(b) Bubble diagram including the vertex correction.
(c) Bethe-Salpeter equation for the vertex corrections $\Lambda$ from electron-phonon interactions within the ladder approximation. The wavy line is the phonon propagator and the red dots are the $e$-ph matrix elements $g_{nm\nu}(\kk,\qq)$. 
}\label{fig:bubble_bse}
\end{figure}
\subsection{Kubo formula and correlation function}
\vspace{-8pt}
We consider a complex vector operator $\mathbf{\hat{A}}$, with matrix elements in the direction $\alpha$ written as $A^\alpha_{nm\kk}=\bra{m\kk} \hat{A}^\alpha \ket{n\kk}$. 
We derive the $\mathbf{\hat{A}}-\mathbf{\hat{A}}$ correlation function with a procedure analogous to the derivation of the dc conductivity in the ladder approximation~\cite{kimVertex2019}. 
Here, the operator $\mathbf{\hat{A}}$ is in general non-diagonal in the band index, leading to matrix elements $A^\alpha_{nm\kk}$, so the derivation for the diagonal case given in Ref.~\cite{kimVertex2019} needs to be extended to non-diagonal operators and vertex corrections. 
\\
\indent
We first derive the correlation function in imaginary time and frequency, and then extend it to real frequencies via analytic continuation. %
The retarded correlation function for the operator $\mathbf{\hat{A}}$ can be obtained from the Kubo formula~\cite{mahanManyParticle2000}
\begin{equation}
    \chi_{\alpha\beta}(\pp,i\nu_b)=\int_0^\beta d\tau e^{i\nu_b\tau}\left< T_\tau \hat{A}^\alpha(\pp,\tau) \hat{A}^\beta(-\pp,0) \right>,
\end{equation}
where $\pp$ is a wave-vector, $\nu_b$ is a bosonic Matsubara frequency, $\tau$ is imaginary time ranging from $0$ to $\beta=1/{k_B T}$ at temperature $T$, and $T_\tau$ is the imaginary time-ordering operator.  %
Here we focus on the $\pp\rightarrow0$ limit, so we drop $\pp$ from the equations.
This correlation function can be expressed as a sum of bubble diagrams $P$ as~\cite{mahanManyParticle2000} 
\begin{equation}
     \chi_{\alpha\beta}(i\nu_b)=\frac{1}{\beta}\sum_{i\omega_a} P(i\omega_a, i\omega_a+i\nu_b).
\end{equation}
Let us consider the bare bubble diagram that includes the electron self-energy only in the electron propagator $\mathcal{G}$, as shown in Fig.~\ref{fig:bubble_bse}(a):
\begin{equation}\label{eq:bubble_bare}
    \chi_{\alpha\beta}(i\nu_b)=\frac{1}{\beta V_{\text{uc}}}\! \sum_{i\omega_a}\Tr \big[ \mathcal{G} (i\omega_a)\hat{A}^\alpha
    \mathcal{G}(i\omega_a+i\nu_b)\hat{A}^\beta \big],
\end{equation}
where the trace is evaluated over the band and momentum indices. 
In this expression, the operator $\mathbf{\hat{A}}$ can be regarded as the bare vertex of the correlation function.
For the velocity operator, Eq.~(\ref{eq:bubble_bare}) leads to the well-known Drude conductivity~\cite{kimVertex2019,mahanManyParticle2000}.
\\
\indent
In this work, the corrections to the vertex originate from the $e$-ph interactions, which couple electronic states with different band and crystal momenta. Figure~\ref{fig:bubble_bse}(b) shows the correlation function including the vertex correction $\Lambda$,
\begin{equation}\label{eq:bubble_ladder}
\begin{split}
    \chi_{\alpha\beta}(i\nu_b)=\frac{1}{\beta V_{\text{uc}}}\! \sum_{i\omega_a}  \Tr & \big[ \mathcal{G} (i\omega_a)\hat{A}^\alpha
    \mathcal{G}(i\omega_a+i\nu_b)\\
    &\times \hat{A}^\beta \Lambda^\beta(i\omega_a,i\omega_a+i\nu_b) \big],
\end{split}
\end{equation}
where $\hat{A}^\beta \Lambda^\beta(i\omega_a,i\omega_a+i\nu_b)$ is the phonon-dressed vertex for the operator $\hat{A}$ in the Cartesian direction $\beta$.
Note that the vertex correction ${\Lambda}^\beta(i\omega_a,i\omega_a+i\nu_b)$ is a complex-valued vector that contains information about the operator dynamics renormalized by the $e$-ph interactions.
\subsection{Bethe-Salpeter equation for the phonon-dressed vertex}
\vspace{-10pt}
The leading correction to the vertex is obtained by summing over ladder diagrams, which can be viewed as an abstract form of charge conservation in the presence of $e$-ph scattering~\cite{mahanManyParticle2000,kimVertex2019}. 
The vertex correction $\Lambda^\alpha_{nn'\kk}$ satisfies the self-consistent BSE, shown diagrammatically in Fig.~\ref{fig:bubble_bse}(c) and written as
\begin{equation} \label{eq:bse_qc}
\begin{split}
A^\alpha_{nn'\kk}&\Lambda^\alpha_{nn'\kk}(i\omega_a,i\omega_a+i\nu_b)=A^\alpha_{nn'\kk}\\
-&\frac{1}{\beta N_q V_{\text{uc}}}\sum_{mm'll'\qq\nu,iq_c} \left[g_{n'm'\nu}(\kk,\qq)\right]^* g_{nm\nu}(\kk,\qq)\mathcal{D}_{\nu\qq}(iq_c)\\
&\times \mathcal{G}_{ml\kk+\qq}(i\omega_a+iq_c) \mathcal{G}_{l'm'\kk+\qq}(i\omega_a+i\nu_b+iq_c) \\
&\times A^\alpha_{ll'\kk} \Lambda^\alpha_{ll'\kk+\qq}(i\omega_a+iq_c,i\omega_a+i\nu_b+iq_c).
\end{split}
\end{equation}
The kernel of this BSE~\cite{stefanucciNonequilibrium2013} is the $e$-ph interaction $\left[g_{n'm'\nu}(\kk,\qq)\right]^* g_{nm\nu}(\kk,\qq)\mathcal{D}_{\nu\qq}(iq_c)$.
\\
\indent
Following Mahan~\cite{mahanManyParticle2000} and Ref.~\cite{kimVertex2019}, we first sum over the bosonic Matsubara frequency $iq_c$ in Eq.~(\ref{eq:bse_qc}). This summation, defined as $S(i\omega_a,i\omega_a+i\nu_b)$, reads:
\begin{equation} \label{eq:bse_contour}
\begin{split}
S&(i\omega_a,i\omega_a+i\nu_b)=\sum_{ll'}S_{ll'}(i\omega_a,i\omega_a+i\nu_b)\\
=&\frac{1}{\beta}\sum_{ll' iq_c}\, \mathcal{D}_{\nu\qq}(iq_c)\,\Lambda^\alpha_{ll'\kk+\qq}(i\omega_a+iq_c,i\omega_a+i\nu_b+iq_c)\\
&\times \mathcal{G}_{ml\kk+\qq}(i\omega_a+iq_c)\,\mathcal{G}_{l'm'\kk+\qq}(i\omega_a+i\nu_b+iq_c).
\end{split}
\end{equation}
As usual, the summation is done by constructing a contour integral along a circle at infinity:
\begin{equation} \label{eq:contour_full}
\begin{split}
\oint \frac{dz}{2 \pi i}& n_{\rm B}(z) \mathcal{D}_{\nu\qq}(z)\, \Lambda^\alpha_{ll'\kk+\qq}(i\omega_a+z,i\omega_a+i\nu_b+z)\\ \times\, &\mathcal{G}_{ml\kk+\qq}(i\omega_a+z)\, \mathcal{G}_{l'm'\kk+\qq}(i\omega_a+i\nu_b+z),
\end{split}
\end{equation}
where $n_{\rm B}$ are Bose-Einstein occupations. 
The integrand has poles at $z=iq_c$, $z=\pm \omega_{\nu\qq}$, and branch cuts along $z=-i\omega_a$ and $z=-i\omega_a-i\nu_b$~\cite{mahanManyParticle2000,kimVertex2019}. 
Employing Cauchy's residue theorem, we obtain 
\begin{equation} \label{eq:bse_qcsum}
\begin{split}
S_{ll'}(i&\omega_a,i\omega_a+i\nu_b)\\
=-&N_{\nu\qq} \Lambda^\alpha_{ll'\kk+\qq}(i\omega_a+\omega_{\nu\qq},i\omega_a+i\nu_b+\omega_{\nu\qq})\\
&\times \mathcal{G}_{ml\kk+\qq}(i\omega_a+\omega_{\nu\qq})\mathcal{G}_{l'm'\kk+\qq}(i\omega_a+i\nu_b+\omega_{\nu\qq})\\
-&[N_{\nu\qq}+1] \Lambda^\alpha_{ll'\kk+\qq}(i\omega_a-\omega_{\nu\qq},i\omega_a+i\nu_b-\omega_{\nu\qq})\\
&\times \mathcal{G}_{ml\kk+\qq}(i\omega_a-\omega_{\nu\qq})\mathcal{G}_{l'm'\kk+\qq}(i\omega_a+i\nu_b-\omega_{\nu\qq})\\
-&\int\frac{d\varepsilon'}{2\pi i} f(\varepsilon')\frac{2\omega_{\nu\qq}}{(\varepsilon'-i\omega_a)^2-\omega_{\nu\qq}^2}\mathcal{G}_{l'm'\kk+\qq}(\varepsilon'+i\nu_b)\\
&\times 
[\Lambda^\alpha_{ll'\kk+\qq}(\varepsilon'+i\eta, \varepsilon'+i\nu_b)\mathcal{G}_{ml\kk+\qq}(\varepsilon'+i\eta)\\
&-\Lambda^\alpha_{ll'\kk+\qq}(\varepsilon'-i\eta, \varepsilon'+i\nu_b)\mathcal{G}_{ml\kk+\qq}(\varepsilon'-i\eta)]\\
-&\int\frac{d\varepsilon'}{2\pi i} f(\varepsilon')\frac{2\omega_{\nu\qq}}{(\varepsilon'-i\omega_a-i\nu_b)^2-\omega_{\nu\qq}^2}\mathcal{G}_{ml\kk+\qq}(\varepsilon'-i\nu_b)\\
&\times 
[\Lambda^\alpha_{ll'\kk+\qq}(\varepsilon'-i\nu_b, \varepsilon'+i\eta)\mathcal{G}_{l'm'\kk+\qq}(\varepsilon'+i\eta)\\
&-\Lambda^\alpha_{ll'\kk+\qq}(\varepsilon'-i\nu_b, \varepsilon'-i\eta)\mathcal{G}_{l'm'\kk+\qq}(\varepsilon'-i\eta)]\,,
\end{split}
\end{equation}
where $N_{\nu \qq} \!=\! n_{\rm B}(\omega_{\nu \qq})$ are temperature dependent phonon occupations, $f(\varepsilon)$ is the Fermi-Dirac distribution function, and $\eta$ is a positive infinitesimal.
\\
\indent
The leading contribution to $S_{ll'}(i\omega_a,i\omega_a+i\nu_b)$ comes from the combination of retarded and advanced Green's functions, $G^R$ and $G^A$, while terms of $O([G^R]^2,[G^A]^2)$ can be neglected at low electron density~\cite{mahanManyParticle2000,kimVertex2019}. 
Therefore, after the analytic continuations $i\omega_a \rightarrow \varepsilon-i\eta$ and $i\omega_a+i\nu_b \rightarrow \varepsilon+\nu+i\eta$, and using the identity $\frac{1}{x+i\eta}=P\frac{1}{x}-i \pi \delta(x)$, we obtain $S_{ll'}(\varepsilon-i\eta,\varepsilon+i\eta)$ in limit of $\nu\rightarrow 0$, %
\begin{equation} \label{eq:SRA}
\begin{split}
S_{ll'}(\varepsilon-&i\eta,\varepsilon+i\eta)\\
=-[&N_{\nu\qq}+f(\varepsilon+\omega_{\nu\qq})]\Lambda^{\alpha}_{ll'\kk+\qq}(\varepsilon+\omega_{\nu\qq})\\
&\times G_{ml\kk+\qq}^R(\varepsilon+\omega_{\nu\qq})G_{l'm'\kk+\qq}^A(\varepsilon+\omega_{\nu\qq})\\
-[&N_{\nu\qq}+1-f(\varepsilon-\omega_{\nu\qq})]\Lambda^{\alpha}_{ll'\kk+\qq}(\varepsilon-\omega_{\nu\qq})\\
&\times G_{ml\kk+\qq}^R(\varepsilon-\omega_{\nu\qq})G_{l'm'\kk+\qq}^A(\varepsilon-\omega_{\nu\qq}),
\end{split}
\end{equation}
where the index A (R) stands for advanced (retarded) function, and $\Lambda^{\alpha}(\varepsilon) \equiv \Lambda^{\alpha}(\varepsilon-i\eta,\varepsilon+i\eta)$.
\\
\indent
Using this result, we write the self-consistent BSE for the phonon-dressed vertex $\mathbf{\hat{A}}\Lambda$ at energy $\varepsilon$ as:
\begin{equation} \label{eq:bse_full_G}
\begin{split}
&A^\alpha_{nn'\kk}\Lambda^{\alpha}_{nn'\kk}(\varepsilon)=A^\alpha_{nn'\kk}\\
&+\frac{1}{N_q V_{\text{uc}}}\sum_{mm'll'\qq\nu} \left[g_{n'm'\nu}(\kk,\qq)\right]^* g_{nm\nu}(\kk,\qq) A^\alpha_{ll'\kk+\qq} \\
&\times \bigg[ (N_{\nu\qq}+f(\varepsilon+\omega_{\nu\qq}))\Lambda^{\alpha}_{ll'\kk+\qq}(\varepsilon+\omega_{\nu\qq})\\
&\times G_{ml\kk+\qq}^R(\varepsilon+\omega_{\nu\qq})G_{l'm'\kk+\qq}^A(\varepsilon+\omega_{\nu\qq})\\
&+(N_{\nu\qq}+1-f(\varepsilon-\omega_{\nu\qq}))\Lambda^{\alpha}_{ll'\kk+\qq}(\varepsilon-\omega_{\nu\qq})\\
&\times G_{ml\kk+\qq}^R(\varepsilon-\omega_{\nu\qq})G_{l'm'\kk+\qq}^A(\varepsilon-\omega_{\nu\qq}) \bigg].
\end{split}
\end{equation}
By solving Eq.~(\ref{eq:bse_full_G}), we obtain the phonon-dressed vertex %
$A^\alpha_{nn'\kk}\Lambda^{\alpha}_{nn'\kk}(\varepsilon)$ and its dependence on band, crystal momentum and energy. 
\\
\indent
In the weak scattering regime, where the electron spectral function has a well-defined quasiparticle peak~\cite{zhouPredicting2019} and the off-diagonal self-energy can be neglected~\cite{lihmPhononinduced2020,allenTheory1976}, the Green's function becomes band-diagonal and the self-energies can be evaluated on-shell.
Then the product of the retarded and advanced Green's functions, $G^R G^A$, can be approximated as~\cite{vollhardtDiagrammatic1980} %
\begin{equation}\label{eq:GRGA}
\begin{split}
    &G^R_{m\kk+\qq}(\varepsilon)G^A_{m'\kk+\qq}(\varepsilon) \\
    &= \frac{G_{m'\kk+\qq}^A(\varepsilon)-G_{m\kk+\qq}^R(\varepsilon)}{G_{m\kk+\qq}^R(\varepsilon)^{-1}-G_{m'\kk+\qq}^A(\varepsilon)^{-1}}\\
    &\approx \frac{\pi\delta(\varepsilon\!-\!\varepsilon_{m'\kk+\qq})\!+\!\pi\delta(\varepsilon\!-\!\varepsilon_{m\kk+\qq})\!-\!iP\frac{1}{\varepsilon-\varepsilon_{m'\kk+\qq}}\!+\!iP\frac{1}{\varepsilon-\varepsilon_{m\kk+\qq}}}{i(\Sigma^R_{m\kk+\qq}-\Sigma^A_{m'\kk+\qq})+i(\varepsilon_{m\kk+\qq}-\varepsilon_{m'\kk+\qq})},
\end{split}
\end{equation}
a function that is strongly peaked at electron energies $\varepsilon\!=\!\varepsilon_{m\kk+\qq}$ and $\varepsilon\!=\!\varepsilon_{m'\kk+\qq}$.
Therefore, we can further simplify the full-frequency BSE in Eq.~(\ref{eq:bse_full_G}) to a double-pole ansatz, which evaluates the vertex corrections only at these two energies:
\begin{widetext}
\begin{equation} \label{eq:bse}
\begin{split}
&A^\alpha_{nn'\kk}\Lambda^{\alpha}_{nn'\kk}(\varepsilon)=A^\alpha_{nn'\kk}+\frac{2\pi}{N_q V_\text{uc}}\sum_{mm'\qq\nu} \left[g_{n'm'\nu}(\kk,\qq)\right]^* g_{nm\nu}(\kk,\qq)  \\
&\times \frac{1}{2} \bigg[ \{ (N_{\nu\qq}+f_{m\kk+\qq})(\delta(\varepsilon+\omega_{\nu\qq}-\varepsilon_{m\kk+\qq})-\frac{i}{\pi}P\frac{1}{\varepsilon+\omega_{\nu\qq}-\varepsilon_{m'\kk+\qq}} )\\
&+(N_{\nu\qq}+1-f_{m\kk+\qq})(\delta(\varepsilon-\omega_{\nu\qq}-\varepsilon_{m\kk+\qq}) -\frac{i}{\pi}P\frac{1}{\varepsilon-\omega_{\nu\qq}-\varepsilon_{m'\kk+\qq}} ) \}\times\frac{A^\alpha_{mm'\kk+\qq}\Lambda^{\alpha}_{mm'\kk+\qq}(\varepsilon_{m\kk+\qq})}{i(\Sigma^R_{m\kk+\qq}-\Sigma^A_{m'\kk+\qq})+i(\varepsilon_{m\kk+\qq}-\varepsilon_{m'\kk+\qq})}\\
&+ \{ (N_{\nu\qq}+f_{m'\kk+\qq})(\delta(\varepsilon+\omega_{\nu\qq}-\varepsilon_{m'\kk+\qq})+\frac{i}{\pi}P\frac{1}{\varepsilon+\omega_{\nu\qq}-\varepsilon_{m\kk+\qq}} )\\
&+(N_{\nu\qq}+1-f_{m'\kk+\qq})(\delta(\varepsilon-\omega_{\nu\qq}-\varepsilon_{m'\kk+\qq}) +\frac{i}{\pi}P\frac{1}{\varepsilon-\omega_{\nu\qq}-\varepsilon_{m\kk+\qq}} ) \}\times\frac{A^\alpha_{mm'\kk+\qq}\Lambda^{\alpha}_{mm'\kk+\qq}(\varepsilon_{m'\kk+\qq})}{i(\Sigma^R_{m\kk+\qq}-\Sigma^A_{m'\kk+\qq})+i(\varepsilon_{m\kk+\qq}-\varepsilon_{m'\kk+\qq})} \bigg],
\end{split}
\end{equation}
\end{widetext}
where $\varepsilon$ equals $\varepsilon_{n\kk}$ or $\varepsilon_{n'\kk}$, and $f_{m\kk+\qq} \equiv f(\varepsilon_{m\kk+\qq})$.
\\
\indent
We have tested the consistency of this theory by deriving a Ward identity~\cite{wardIdentity1950,kimVertex2019,mahanManyParticle2000} relating the self-energy and vertex corrections (see Appendix~\ref{sec:ward}). This result guarantees that $e$-ph diagrams are taken into account consistently in the self-energy and in our BSE.
\subsection{The dressed vertex and its interpretation}
\vspace{-8pt}
\begingroup
\setlength{\tabcolsep}{6pt} %
\renewcommand{\arraystretch}{1.5} %
\begin{table*}[htbp]
\centering
\caption{
Summary of the formalism for charge transport and spin decoherence. 
}
\begin{tabular}{c c c} 
 \hline\hline
  ~ & Charge transport (Ref.~\cite{kimVertex2019}) & Spin decoherence \\ [0.5ex] 
 \hline
  Operator & $v_{n\bm{k}}$ (diagonal)  & $s_{nm\bm{k}}$ (non-diagonal) \\ %
  External field $\mathcal{F}$ & Vector potential ($\vec{A}$)     & Magnetic field ($\vec{B}$)   \\
  Injection field $\dot{\mathcal{F}}$ & $\vec{E}(\nu)=-i\nu\vec{A}(\nu)$     & $\dot{\vec{B}}(\nu)=-i\nu\vec{B}(\nu)$   \\
  Vertex correction $\Lambda$ & $\Lambda_{n\kk}^\alpha(\varepsilon_{n\kk})$               & \vspace{2pt}$\Lambda_{nn'\kk}^\alpha(\varepsilon_{n\kk}), \Lambda_{nn'\kk}^\alpha(\varepsilon_{n'\kk})$    \\
  Renormalized dynamics $\tau$, $\omega$ & $\tau_{n\kk}^{(\text{tr})\alpha} = \tau^{\text{e-ph}}_{n\kk}\Lambda_{n\kk}^\alpha(\epsilon_{n\kk})$               &  $\frac{1}{\frac{1}{\tau_{nn'\kk}^\alpha(\varepsilon_{n\kk})}+i{\omega}_{nn'\kk}^\alpha(\varepsilon_{n\kk})} =  \frac{\Lambda^{\alpha}_{nn'\kk}(\varepsilon_{n\kk})}{i(\Sigma^R_{n\kk}-\Sigma^A_{n'\kk}) + i(\varepsilon_{n\kk}-\varepsilon_{n'\kk})}$    \\
  [2.5ex]
 \hline\hline
\end{tabular}
\label{table:summary}
\end{table*}
\endgroup
We focus on the dressed operator divided by the band energy difference, a key term in Eq.~(\ref{eq:bse}):
\begin{equation}\label{eq:full_vertex}
\frac{A^{\alpha}_{mm'\kk+\qq}\Lambda^{\alpha}_{mm'\kk+\qq}(\varepsilon_{m'\kk+\qq})}{i(\Sigma^R_{m\kk+\qq}\!-\!\Sigma^A_{m'\kk+\qq})\!+\!i(\varepsilon_{m\kk+\qq}\!-\!\varepsilon_{m'\kk+\qq})}.
\end{equation} %
This ratio describes the renormalized dynamics associated with the operator $\mathbf{\hat{A}}$ in the presence of $e$-ph interactions.
This dynamics is obtained by dividing Eq.~(\ref{eq:full_vertex}) by the bare operator expectation value $A^\alpha_{mm'\kk+\qq}$, obtaining
\begin{equation}\label{eq:effective_dynamics}
\frac{\Lambda^{\alpha}_{mm'\kk+\qq}(\varepsilon_{m'\kk+\qq})}{i(\Sigma^R_{m\kk+\qq}\!-\!\Sigma^A_{m'\kk+\qq})\!+\!i(\varepsilon_{m\kk+\qq}\!-\!\varepsilon_{m'\kk+\qq})}.
\end{equation}
The physical meaning of this ratio can be understood by analyzing the simple case of the velocity operator. 
As the velocity operator is band-diagonal and satisfies $v^\alpha_{mm'\kk+\qq}=v^\alpha_{m\kk+\qq}\delta_{mm'}$, the band energy difference in the denominator vanishes, so the denominator is purely real because $\Sigma^A_{m'\kk+\qq} = (\Sigma^R_{m'\kk+\qq})^*$. 
Thus Eq.~(\ref{eq:full_vertex}) for the velocity operator becomes
\begin{equation}\label{eq:dressed_mfp}
\frac{v^\alpha_{m\kk+\qq}\Lambda^{\alpha}_{mm\kk+\qq}(\varepsilon_{m\kk+\qq})}{i(\Sigma^R_{m\kk+\qq}-\Sigma^A_{m\kk+\qq})}
= v^\alpha_{m\kk+\qq} \tau^{\text{e-ph}}_{m\kk+\qq} \Lambda^{\alpha}_{mm\kk+\qq}(\varepsilon_{m\kk+\qq}),
\end{equation}
where we used $\tau^{\text{e-ph}}_{m\kk+\qq}=1/|2\Im\Sigma_{m\kk+\qq}|$ for the $e$-ph collision time. %
This equation gives the renormalized $e$-ph mean free path, and dividing 
by the bare velocity we obtain the renormalized relaxation time, also known as the transport relaxation time~\cite{kimVertex2019},
\begin{equation}
\label{eq:tautr}
\tau_{m\kk+\qq}^{\alpha(\text{tr})} \equiv \tau^{\text{e-ph}}_{m\kk+\qq}\, \Lambda^{\alpha}_{mm\kk+\qq}(\varepsilon_{m\kk+\qq}) =  \frac{\Lambda^{\alpha}_{mm\kk+\qq}(\varepsilon_{m\kk+\qq})}{{i(\Sigma^R_{m\kk+\qq}-\Sigma^A_{m\kk+\qq}})}.
\end{equation}
\\
\indent
For a non-diagonal operator, both the vertex correction and the operator expectation value are complex, so the ratio in Eq.~(\ref{eq:effective_dynamics}) cannot be represented by a single real quantity with units of time as in Eq.~(\ref{eq:tautr}).  
To extend the vertex correction to non-diagonal operators, we generalize this formalism by defining the renormalized microscopic relaxation times $\tau_{mm'\kk+\qq}^\alpha(\varepsilon)$ and introducing the precession frequencies $\omega_{mm'\kk+\qq}^\alpha(\varepsilon)$:
\begin{equation}\label{eq:dressed_time}
\begin{split}
&\frac{1}{\frac{1}{\tau_{mm'\kk+\qq}^\alpha(\varepsilon)}+i{\omega}_{mm'\kk+\qq}^\alpha(\varepsilon) }\\
&\equiv
\frac{\Lambda^{\alpha}_{mm'\kk+\qq}(\varepsilon)}{i(\Sigma^R_{m\kk+\qq}\!-\!\Sigma^A_{m'\kk+\qq})\!+\!i(\varepsilon_{m\kk+\qq}\!-\!\varepsilon_{m'\kk+\qq})},
\end{split}
\end{equation}
where $\varepsilon$ equals $\varepsilon_{m\kk+\qq}$ or $\varepsilon_{m'\kk+\qq}$. 
This way, without the vertex correction, the renormalized relaxation time reduces to the (non-diagonal) $e$-ph collision time,   $\tau^{\text{e-ph}}_{mm'\kk+\qq}=1/|\Im\Sigma_{m\kk+\qq}+\Im\Sigma_{m'\kk+\qq}|$, and the renormalized precession frequency reduces to the bare operator rotation frequency, $\omega_{\textrm{B}} = (\varepsilon_{m\kk+\qq}+\Re\Sigma_{m\kk+\qq})-(\varepsilon_{m'\kk+\qq}+\Re\Sigma_{m'\kk+\qq})$, with $A^{\alpha}_{mm'\kk+\qq}(t) \propto e^{i\, {\omega}_{\textrm{B}} t}$. 
\subsection{Vertex correction to the susceptibility} %
\vspace{-8pt}
We derive the vertex-corrected susceptibility in response to an external field for the generic observable $\mathbf{\hat{A}}$. 
Suppose that the complex operator $\hat{A}^\alpha$ couples to a vector field $\mathcal{F}^\alpha$, with perturbation Hamiltonian $H' = -\hat{A}^\alpha \mathcal{F}^\alpha$.
The susceptibility is defined as the response function in
\begin{equation}\label{eq:m_susceptibility}
    \langle\hat{A}^\alpha(\nu)\rangle=\chi_{\alpha\beta}(\nu)\mathcal{F}^\beta(\nu),
\end{equation}
where $\mathcal{F}$ is the external field along the direction $\beta$, and $\langle \hat{A}^\alpha(\nu) \rangle$ is the response of the system along $\alpha$ at frequency $\nu$ due to the applied field. 
\\
\indent
To study relaxation and dissipation, we rewrite the response of the system as 
\begin{equation}\label{eq:s_injection}
    \langle\hat{A}^\alpha(\nu)\rangle=\sigma_{\alpha\beta}(\nu)\dot{\mathcal{F}}^\beta(\nu),
\end{equation}
thus expressing it in terms of the susceptibility $\sigma_{\alpha\beta}$ to the ``injection field'' at frequency $\nu$, and
$\dot{\mathcal{F}}^\beta(\nu)=-i\nu \mathcal{F}^\beta(\nu)$.
The injection field produces a nonequilibrium electron distribution with an injection rate equal to the inverse relaxation time of $\mathbf{\hat{A}}$~\cite{shenMicroscopic2014}.
From Eqs.~(\ref{eq:m_susceptibility})-(\ref{eq:s_injection}), we obtain   
\begin{equation}
    \sigma_{\alpha\beta}(\nu)=\frac{\chi_{\alpha\beta}(\nu)}{-i\nu}.
\end{equation}
\\
\indent
When $\mathcal{F}$ is the vector potential $\mathbf{A}$, the injection field becomes the electric field $\mathbf{E}(\nu)=-i\nu\mathbf{A}(\nu)$, the observable of interest is the current operator  $A^\alpha_{nm\kk}= e \delta_{nm}v^\alpha_{n\kk}$, and $\sigma_{\alpha\beta}(\nu)$ is the frequency-dependent conductivity tensor. When $\mathcal{F}$ is the magnetic field $\mathbf{B}$,
the injection field is its time derivative,  $\dot{\mathbf{B}}(\nu)=-i\nu\mathbf{B}(\nu)$, and the observable is the electron magnetic moment $A^\alpha_{nm\kk}=g \mu_B s^\alpha_{nm\kk}$, which is proportional to the spin matrix 
$s^\alpha_{nm\kk}$~\cite{shenMicroscopic2014,offidaniMicroscopic2018}. 
These results are summarized in Table~\ref{table:summary}.
\\
\indent
We write the correlation function with vertex correction [see Eq.~(\ref{eq:bubble_ladder})] as a contour integral along a circle at infinity~\cite{mahanManyParticle2000,kimVertex2019},
\begin{equation}
\begin{split}
\chi_{\alpha\beta}(i\nu_b)=&-\frac{1}{V_{\text{uc}}}\oint \frac{dz}{2\pi i} f(z) \Tr \big[ \\ &
\mathcal{G} (z)\hat{A}^\alpha\mathcal{G}(z+i\nu_b) \hat{A}^\beta \Lambda^\beta(z,z+i\nu_b) \big],
\end{split}
\end{equation}
which has branch cuts along $z=-i\nu_b$ and $z=0$, and poles at $z=i\omega_a$, and thus 
\begin{equation}
\begin{split}
&\chi_{\alpha\beta}(i\nu_b)=\frac{1}{V_{\text{uc}}}\int \frac{d\varepsilon}{2\pi i} f(\varepsilon) \Tr\bigg[ \\
&- \mathcal{G} (\varepsilon+i\eta)\hat{A}^\alpha \mathcal{G}(\varepsilon+i\nu_b) \hat{A}^\beta \Lambda^\beta(\varepsilon+i\eta,\varepsilon+i\nu_b) \\
&+ \mathcal{G} (\varepsilon-i\eta)\hat{A}^\alpha \mathcal{G}(\varepsilon+i\nu_b) \hat{A}^\beta \Lambda^\beta(\varepsilon-i\eta,\varepsilon+i\nu_b) \\
&- \mathcal{G} (\varepsilon-i\nu_b)\hat{A}^\alpha \mathcal{G}(\varepsilon+i\eta) \hat{A}^\beta \Lambda^\beta(\varepsilon-i\nu_b,\varepsilon+i\eta) \\
&+ \mathcal{G} (\varepsilon-i\nu_b)\hat{A}^\alpha \mathcal{G}(\varepsilon-i\eta) \hat{A}^\beta \Lambda^\beta(\varepsilon-i\nu_b,\varepsilon-i\eta)  \bigg].
\end{split}
\end{equation}
After the analytic continuation $i\nu_b\rightarrow\nu+i\eta$, we obtain the retarded correlation function to leading order by neglecting the terms $G^R G^R$ and $G^A G^A$~\cite{mahanManyParticle2000,kimVertex2019}:
\begin{equation}\label{eq:s-s-corr}
\begin{split}
    &\chi_{\alpha\beta}(\nu)\\
    &=\frac{1}{V_{\text{uc}}}\int\frac{d\varepsilon}{2\pi i}(f(\varepsilon)-f(\varepsilon+\nu))\Tr\big[  \\
    &~~~~G^R (\varepsilon)\hat{A}^\alpha G^A(\varepsilon+\nu) \hat{A}^\beta \Lambda^\beta(i\omega_a-i\eta,i\omega_a+i\nu_b+i\eta) \big]\\
    &\approx \frac{1}{N_k V_\text{uc}} \sum_{nm\kk} \int\frac{d\varepsilon}{2\pi i}(f(\varepsilon)-f(\varepsilon+\nu))A^\alpha_{nm\kk}A^\beta_{mn\kk}\\
    &\times \!\Lambda^{\beta}_{mn\kk}(\varepsilon\!-\!i\eta,\varepsilon\!+\!\nu\!+\!i\eta)\frac{\pi\delta(\varepsilon\!-\!\varepsilon_{n\kk})\!+\!\pi\delta(\varepsilon\!+\!\nu\!-\!\varepsilon_{m\kk})}{i(\Sigma^R_{m\kk}\!-\!\Sigma^A_{n\kk})\!+\!i(\varepsilon_{m\kk}\!+\!\nu\!-\!\varepsilon_{n\kk})},
\end{split}
\end{equation}
where $N_k$ is the number of $\kk$-points, and we used Eq.~(\ref{eq:GRGA}) in the last equality.  
This equation characterizes the frequency-dependent response of the system~\cite{colemanIntroduction2015}.
\\
\indent
We focus on the dc limit $\nu\rightarrow0$, where the driving field is static. 
The susceptibility with respect to the injection field becomes
\begin{equation}\label{eq:s-sus-inj}
\begin{split}
    &\lim_{\nu\rightarrow 0}\sigma_{\alpha\beta}(\nu)=-\lim_{\nu\rightarrow 0} \frac{1}{\nu}\Im\chi_{\alpha\beta}(\nu)\\
    &=\frac{1}{N_k V_{\text{uc}}} \Re \sum_{nm\kk}A^\alpha_{nm\kk}A^\beta_{mn\kk} \\
    &~~~~~\times\frac{\frac{1}{2}[(-\frac{df_{n\kk}}{d\varepsilon})\Lambda^\beta_{mn\kk}(\varepsilon_{n\kk})+(-\frac{df_{m\kk}}{d\varepsilon})\Lambda^\beta_{mn\kk}(\varepsilon_{m\kk})]}{{i(\Sigma^R_{m\kk}-\Sigma^A_{n\kk})+i(\varepsilon_{m\kk}-\varepsilon_{n\kk})}}.%
\end{split}
\end{equation}
\\
\indent %
This static susceptibility has both band-diagonal ($n=m$) and off-diagonal ($n\ne m$) contributions.  
The band-diagonal contribution %
\begin{equation}\label{eq:s-sus-inj-diag}
    \sigma_{\alpha\beta}^{\rm (d)}(0) = \frac{1}{N_k V_{\text{uc}}} \sum_{n\kk} A^\alpha_{nn\kk}A^\beta_{nn\kk}\tau^\text{e-ph}_{n\kk}\Lambda^{\beta}_{nn\kk}(\varepsilon_{n\kk}) (-\frac{df_{n\kk}}{d\varepsilon})
\end{equation} %
is the only contribution to the static susceptibility for a band-diagonal operator. For example, for the velocity operator, Eq.~(\ref{eq:s-sus-inj-diag}) becomes the well-known electrical conductivity tensor within the BTE, which has been studied extensively using both empirical and first-principles calculations~\cite{Ziman, Lundstrom, liElectrical2015,zhouInitio2016,zhouPerturbo2021} (note that solving exactly the BTE is equivalent to computing the velocity vertex correction~\cite{kimVertex2019}). 
For non-diagonal operators, our formalism introduces an off-diagonal contribution in Eq.~(\ref{eq:s-sus-inj}):
\begin{equation}\label{eq:s-sus-inj-off}
\begin{split}
    &\sigma_{\alpha\beta}^{\rm (nd)}(0) = \frac{1}{N_{\kk} V_{\text{uc}}} \Re \sum_{n\ne m\kk}A^\alpha_{nm\kk}A^\beta_{mn\kk} \\
    &~~~~~\times\frac{\frac{1}{2}[(-\frac{df_{n\kk}}{d\varepsilon})\Lambda^\beta_{m n\kk}(\varepsilon_{n\kk})+(-\frac{df_{m\kk}}{d\varepsilon})\Lambda^\beta_{mn\kk}(\varepsilon_{m\kk})]}{{i(\Sigma^R_{m\kk}-\Sigma^A_{n\kk})+i(\varepsilon_{m\kk}-\varepsilon_{n\kk})}}.
\end{split}
\end{equation}
For the velocity operator, this term enables studies of charge transport in the presence of inter-band coherence~\cite{culcerInterband2017}; for the spin operator, this contribution is essential to describe how $e$-ph interactions modify spin precession.\\
\subsection{Renormalized relaxation time}
\vspace{-8pt}
We derive an expression for the renormalized macroscopic relaxation time of an operator $\mathbf{\hat{A}}$ due to $e$-ph interactions. Using Eq.~(\ref{eq:s-sus-inj-diag}), the average relaxation time for the band-diagonal components $A^\alpha_{nn\kk}$ is
\begin{equation}\label{eq:srt_def}
\begin{split}
    \tau_{\alpha\beta} %
    =\frac{\sum_{n\kk} A^\alpha_{nn\kk}A^\beta_{nn\kk}\tau^\text{e-ph}_{n\kk}\Lambda^{\beta}_{nn\kk}(\varepsilon_{n\kk}) (-\frac{df_{n\kk}}{d\varepsilon})}{\sum_{n\kk}   A^\alpha_{nn\kk}A^\beta_{nn\kk}
    \big(-\frac{df_{n\kk}}{d\varepsilon} \big)}.
\end{split} 
\end{equation}
For the velocity operator, this equation gives the well-known Drude dc electrical conductivity, while for the spin operator one obtains the phonon-dressed macroscopic spin relaxation time, as discussed below. 
These results generalize the linear response treatment for band-diagonal operators presented in Ref.~\cite{kimVertex2019} and extend it to non-diagonal operators.\\

\section{Spin relaxation and decoherence}\label{sec:spin}
We now specialize to the non-diagonal spin operator, and apply our formalism to study phonon-induced spin relaxation and decoherence. The BSE for the phonon-dressed spin vertex $-$ called hereafter spin-phonon BSE $-$ is a key result obtained from Eq.~(\ref{eq:bse_full_G}) by replacing $A^\alpha_{nn'\kk}$ with the spin operator $s^\alpha_{nn'\kk}$. In matrix form and using a compact notation, the spin-phonon BSE can be written in a way that clearly matches the diagram in Fig.~\ref{fig:bubble_bse}(c):
\begin{equation} \label{eq:bse_matrix}
\begin{split}
    \bm{s}\bm{\Lambda}_{\kk}(\varepsilon)\!=\!\bm{s}_{\kk}+\frac{1}{N_q V_\text{uc}}\!\sum_{\nu\qq\pm} 
    &\textbf{g}_{\nu\kk\qq}^\dagger 
    \!\left[{G}^A\bm{s}\bm{\Lambda} 
     {G}^R \right]_{\!\!\begin{smallmatrix}\kk+\qq,~~ \\ \varepsilon\pm\omega_{\nu\qq}\end{smallmatrix}} 
     \!\!\!\textbf{g}_{\nu\kk\qq}\, F_{\pm}(T),
\end{split}
\end{equation} 
where $\bm{s}\bm{\Lambda}_{\kk}(\varepsilon) = \bm{s}_{nn'\kk} \bm{\Lambda}_{nn'\kk}(\varepsilon)$ is the phonon-dressed spin vertex, $F_\pm(T)=N_{\nu\qq}+\frac{1}{2}\pm[f(\varepsilon\pm\omega_{\nu\qq})-\frac{1}{2}]$ is a thermal occupation factor at temperature $T$, and $\left[\mathbf{g}_{\nu\kk\qq}\right]_{nm}=g_{nm\nu}(\kk,\qq)$ are $e$-ph matrix elements~\cite{zhouPerturbo2021}.
\\
\indent
In the weak scattering regime, this spin-phonon BSE can be rewritten using the double-pole ansatz discussed above (where $\varepsilon$ equals $\varepsilon_{n\kk}$ or $\varepsilon_{n'\kk}$):
\begin{widetext}
\begin{equation} \label{eq:bse_spin}
\begin{split}
&s^\alpha_{nn'\kk}\Lambda^{\alpha}_{nn'\kk}(\varepsilon)=s^\alpha_{nn'\kk}+\frac{2\pi}{N_q V_\text{uc}}\sum_{mm'\qq\nu} \left[g_{n'm'\nu}(\kk,\qq)\right]^* g_{nm\nu}(\kk,\qq)  \\
&\times \frac{1}{2} \bigg[ \{ (N_{\nu\qq}+f_{m\kk+\qq})(\delta(\varepsilon+\omega_{\nu\qq}-\varepsilon_{m\kk+\qq})-\frac{i}{\pi}P\frac{1}{\varepsilon+\omega_{\nu\qq}-\varepsilon_{m'\kk+\qq}} )\\
&+(N_{\nu\qq}+1-f_{m\kk+\qq})(\delta(\varepsilon-\omega_{\nu\qq}-\varepsilon_{m\kk+\qq}) -\frac{i}{\pi}P\frac{1}{\varepsilon-\omega_{\nu\qq}-\varepsilon_{m'\kk+\qq}} ) \}\times\frac{s^\alpha_{mm'\kk+\qq}\Lambda^{\alpha}_{mm'\kk+\qq}(\varepsilon_{m\kk+\qq})}{i(\Sigma^R_{m\kk+\qq}-\Sigma^A_{m'\kk+\qq})+i(\varepsilon_{m\kk+\qq}-\varepsilon_{m'\kk+\qq})}\\
&+ \{ (N_{\nu\qq}+f_{m'\kk+\qq})(\delta(\varepsilon+\omega_{\nu\qq}-\varepsilon_{m'\kk+\qq})+\frac{i}{\pi}P\frac{1}{\varepsilon+\omega_{\nu\qq}-\varepsilon_{m\kk+\qq}} )\\
&+(N_{\nu\qq}+1-f_{m'\kk+\qq})(\delta(\varepsilon-\omega_{\nu\qq}-\varepsilon_{m'\kk+\qq}) +\frac{i}{\pi}P\frac{1}{\varepsilon-\omega_{\nu\qq}-\varepsilon_{m\kk+\qq}} ) \}\times\frac{s^\alpha_{mm'\kk+\qq}\Lambda^{\alpha}_{mm'\kk+\qq}(\varepsilon_{m'\kk+\qq})}{i(\Sigma^R_{m\kk+\qq}-\Sigma^A_{m'\kk+\qq})+i(\varepsilon_{m\kk+\qq}-\varepsilon_{m'\kk+\qq})} \bigg].
\end{split}
\end{equation}
\end{widetext}
This BSE for the phonon-dressed spin vertex, used in this work to study spin dynamics, should not be confused with the widely used BSE for excitons and optical spectra~\cite{rohlfingElectronhole2000}, which is entirely unrelated. 
\\
\indent
The vertex corrections $\Lambda^{\alpha}_{nn'\kk}$ obtained by solving the BSE govern spin dynamics as they renormalize spin relaxation and precession~\cite{parkPredicting2022}.  
The \textit{macroscopic} spin relaxation times are obtained using the thermal average in Eq.~(\ref{eq:srt_def}),
\begin{equation}\label{eq:srt}
    \tau^{(s)}_{\alpha\beta}=\frac{\sum_{n\kk} s^\alpha_{nn\kk}s^\beta_{nn\kk}\tau^\text{e-ph}_{n\kk}\Lambda^{\beta}_{nn\kk}(\varepsilon_{n\kk}) (-\frac{df_{n\kk}}{d\varepsilon})}{\sum_{n\kk}   s^\alpha_{nn\kk}s^\beta_{nn\kk}
    \,\big(\!-\frac{df_{n\kk}}{d\varepsilon} \big)}.
\end{equation}
For $\alpha=\beta$ along the external magnetic field, Eq.~(\ref{eq:srt}) gives the longitudinal spin relaxation time, usually called $T_1$, along the direction $\alpha$, while for %
a perpendicular magnetic field one obtains the transverse spin relaxation time, $T_2$~\cite{burkovSpin2004}.
The renormalized \textit{microscopic} spin relaxation times ($\tau_{nn'\kk}^\alpha$) and spin precession rates ($\omega_{nn'\kk}^\alpha$), which are matrices in Bloch basis, are computed from the vertex corrections $\Lambda^{\alpha}_{nn'\kk}$ using 
\begin{equation}\label{eq:dressed_time_spin}
    \frac{1}{\frac{1}{\tau_{nn'\kk}^\alpha(\varepsilon)}+i{\omega}_{nn'\kk}^\alpha(\varepsilon)}\equiv\frac{\Lambda^{\alpha}_{nn'\kk}(\varepsilon)}{i(\Sigma^R_{n\kk}-\Sigma^A_{n'\kk})+i(\varepsilon_{n\kk}-\varepsilon_{n'\kk})}.
\end{equation}
The diagonal components with $n\!=\!n'$ give the renormalized microscopic spin relaxation times, $\tau^{\beta}_{nn\kk} = \tau^\text{e-ph}_{n\kk}\, \Lambda^{\beta}_{nn\kk}(\varepsilon_{n\kk})$, entering Eq.~(\ref{eq:srt}). 

\newpage
\section{Results}\label{sec:Results}
We apply our formalism to study spin relaxation and decoherence. We first present analytic results for a two-level model system, and then focus on first-principles calculations on a real material, GaAs. Application to a wider range of materials is presented in our companion paper~\cite{parkPredicting2022}.
\subsection{Two-level system with optical phonon scattering}

\begin{figure}[!t]
\includegraphics[width=0.6\columnwidth]{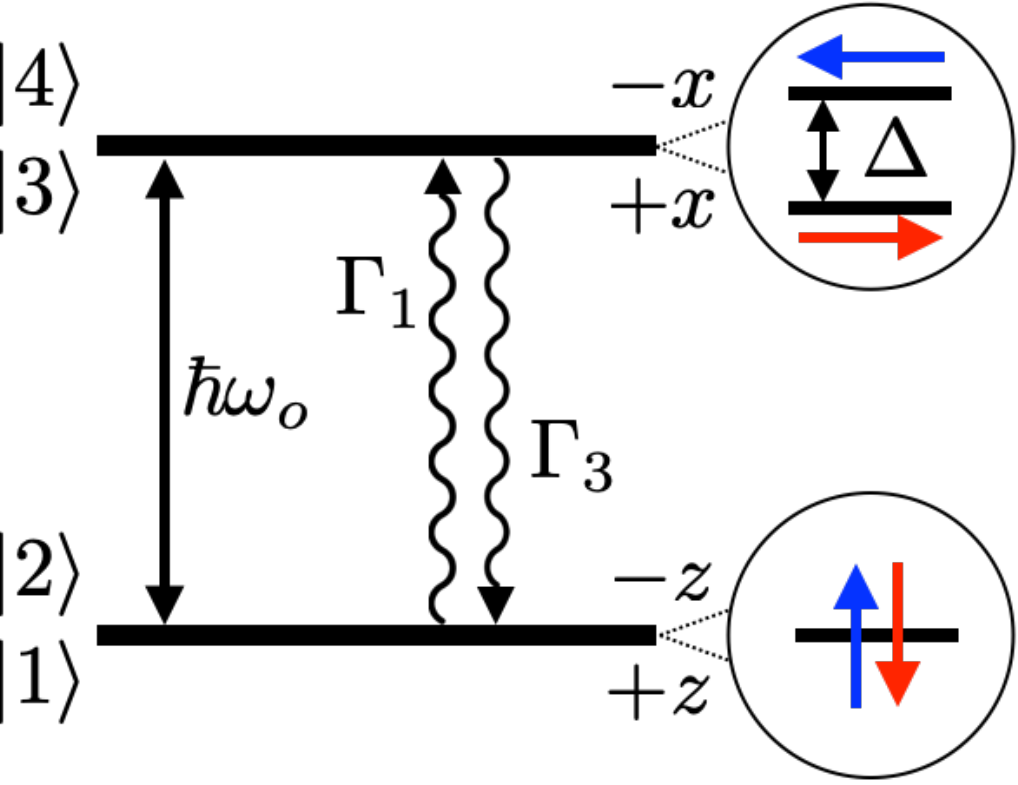}
\caption{Schematic of the energy levels, scattering rates, and spin orientations of the four Bloch states in our model. 
}\label{fig:model_diagram}
\end{figure}

We study spin dynamics in a two-level system to understand different phonon-induced spin relaxation mechanisms. In our model, the electron spins undergo phonon-induced spin-flip transitions together with spin precession in the SOC field modified by the $e$-ph interactions. 
We solve the spin-phonon BSE for this system and derive analytic expressions for the vertex corrections and spin relaxation times. Our analysis sheds light on phonon-dressed operators and their renormalized dynamics, providing a starting point to understand phonon-induced spin relaxation in real materials with complex band structures, phonon dispersions, and $e$-ph interactions.
\\
\indent
Consider a periodic two-level system where each level is spin degenerate (see Fig.~\ref{fig:model_diagram}).
The Hilbert space consists of four Bloch states, which are eigenstates of the Hamiltonian:
\begin{equation}
    H\!\ket{n}=\varepsilon_n \! \ket{n}
\end{equation}
where $\ket{n}$ is the $n$-th energy eigenstate.
The two lowest-energy states $\ket{1}$ and $\ket{2}$ are degenerate ($\varepsilon_1=\varepsilon_2$) and differ only in their spin part.  
The other two states, $\ket{3}$ and $\ket{4}$, are higher in energy by $\omega_O$ and are perturbed by an internal magnetic field along $\hat{x}$ due to SOC. 
This field causes a small Zeeman splitting, $\Delta \ll \omega_O$, such that $\varepsilon_{3,4}=\varepsilon_1+\omega_O \pm \frac{\Delta}{2}$.
\\
\indent
We separate the space-dependent part $\ket{\psi_n}$ and the spin-dependent part $\ket{\chi_n}$ of the two eigenstates as  $\ket{n}=\ket{\psi_n}\otimes\ket{\chi_n}$.
The two lowest states have an identical space-dependent part $\ket{\psi}$ and are spin polarized along $\hat{z}$:
\begin{equation}
    \ket{1}=\ket{\psi}\otimes \left( \begin{matrix} 1 \\ 0 \end{matrix} \right),~ 
    \ket{2}=\ket{\psi}\otimes \left( \begin{matrix} 0 \\ 1 \end{matrix} \right), 
\end{equation}
with the following spin matrix elements along $z$:
\begin{equation}
\begin{split}
    &\bra{1}\hat{s}_z\ket{1}=-\bra{2}\hat{s}_z\ket{2}=\frac{1}{2},\\
    &\bra{1}\hat{s}_z\ket{2}=0.
\end{split}
\end{equation}
Above, $\hat{s}=\hat{\sigma}/2$ is the spin operator and $\hat{\sigma}$ are Pauli matrices. %
The two upper bands have an identical space-dependent part $\ket{\phi}$ and are spin polarized along $\hat{x}$: 
\begin{equation}
    \ket{3}=\ket{\phi}\otimes \frac{1}{\sqrt{2}}\left( \begin{matrix} 1 \\ 1 \end{matrix} \right),~ 
    \ket{4}=\ket{\phi}\otimes \frac{1}{\sqrt{2}}\left( \begin{matrix} 1 \\ -1 \end{matrix} \right), 
\end{equation}
with spin matrix elements
\begin{equation}
\begin{split}
    &\bra{3}\hat{s}_z\ket{3}=\bra{4}\hat{s}_z\ket{4}=0,\\
    &\bra{3}\hat{s}_z\ket{4}=\frac{1}{2}.
\end{split}
\end{equation}
The space-dependent part of the two lower states is orthogonal to that of the upper states, and the spin matrix elements between the two sets of states are zero. 
\\
\indent
In our model, an electron can scatter between the lower and upper levels by emitting or absorbing an optical phonon. These transitions are associated with $e$-ph matrix elements  $g_{nm}=\bra{m}\Delta\hat{ V}\ket{n}$, where $\Delta\hat{V}$ is the perturbation potential due to the optical phonon. 
We assume that this perturbation potential has the form
\begin{equation}\label{eq:dv}
   \Delta \hat{ V}=\Delta \hat{ V}(\bm{r})\otimes
    \left( \begin{matrix} a&b \\ b&a \end{matrix} \right),
\end{equation}
where $\hat{V}(\bm{r})$ is the space-dependent part and $\left( \begin{smallmatrix} a&b \\ b&a \end{smallmatrix} \right)$ the spin-dependent part of the perturbation, with $a$ and $b$ real numbers.
Due to the presence of SOC, the spin-dependent part $\left( \begin{smallmatrix} a&b \\ b&a \end{smallmatrix} \right)$ is different from the identity matrix. 
We consider a small spin-mixing $b$, where $a^2+b^2=1$, so that each phonon collision has a small probability $b^2 \ll 1$ to flip the $z$-component of the spin ~\cite{elliottTheory1954}. 
The $e$-ph matrix elements become %
$g_{nm}=g_0\bra{\chi_m}
\left( \begin{smallmatrix} a&b \\ b&a \end{smallmatrix} \right) \ket{\chi_n}$, 
where $g_0=\bra{\phi}\Delta \hat{V}(\bm{r})\ket{\psi}$. 
Thus the spin-dependent $e$-ph matrix elements are
\begin{equation}
\begin{split}
    g_{13}=g_{23}=\frac{1}{\sqrt{2}}g_0(a+b),\\
    g_{14}=-g_{24}=\frac{1}{\sqrt{2}}g_0(a-b).
\end{split}
\end{equation}
\subsection{Spin-phonon BSE and spin relaxation times}
\begin{figure}[!t]
\includegraphics[width=0.95\columnwidth]{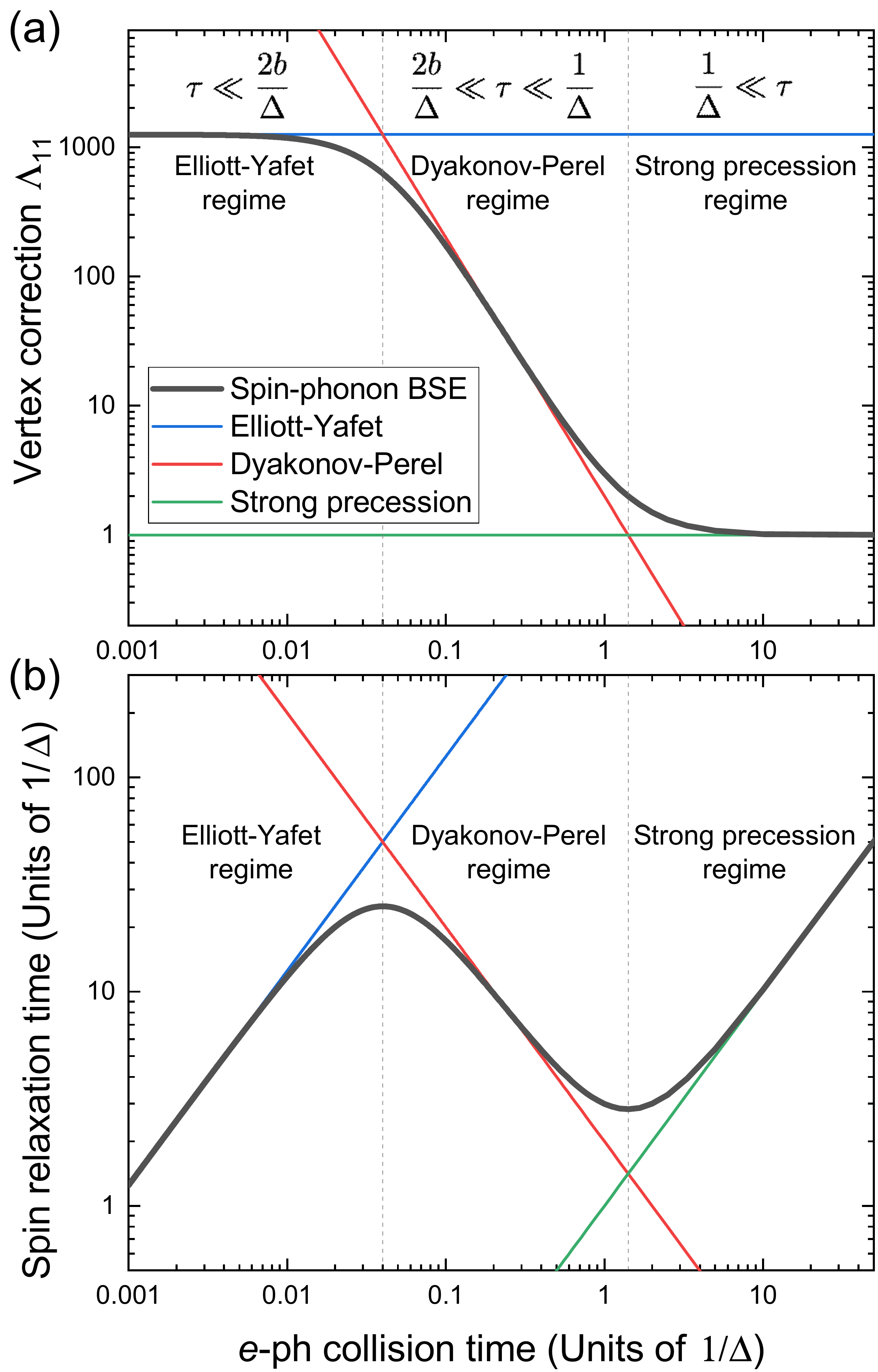}
\caption{(a) The vertex correction $\Lambda^z_{11}$ as a function of the $e$-ph collision time $\tau_1$ for the model system in Fig.~\ref{fig:model_diagram}. (b) Spin relaxation time as a function of the $e$-ph collision time $\tau_1$ for our model system. 
We show the full solution of the spin-phonon BSE in Eqs.~(\ref{eq:Lambda})-(\ref{eq:srt_model})) (black curve) and approximate results for the Elliott-Yafet (blue, Eqs.~(\ref{eq:lambda_ey})-(\ref{eq:srt_ey})), Dyakonov-Perel (red, Eqs.~(\ref{eq:lambda_dp})-(\ref{eq:srt_dp})), and strong-precession regimes (green, Eqs.~(\ref{eq:lambda_strong})-(\ref{eq:srt_opposite_dp})). 
These results are obtained by setting $\tau_3=\tau_1$ and $b=0.02$. 
}\label{fig:srt_result}
\end{figure}

\vspace{-8pt}
We construct the spin-phonon BSE for our two-level system. 
There are four non-zero spin matrix elements in our model: $s^z_{11}$, $s^z_{22}$, $s^z_{34}$, and $s^z_{43}$, where $s^z_{nm}=\bra{m} \hat{s}_z \ket{n}$. 
Using Eq.~(\ref{eq:bse_spin}), the diagonal matrix elements have only one vertex correction, while the off-diagonal matrix elements have two energy-dependent vertex corrections.
Assuming a small Zeeman splitting $\Delta$, we can neglect the energy dependence of the off-diagonal vertex corrections, and define $\Lambda^z_{nm}=\Lambda^{z}_{nm}(\varepsilon_n)$.
We thus have four unknown vertex corrections: $\Lambda^z_{11}$, $\Lambda^z_{22}$, $\Lambda^z_{34}$, and $\Lambda^z_{43}$.
Taking the $e$-ph scattering rates of the two lower and upper states to be $\Gamma_{1}$ and $\Gamma_{3}$,  respectively, the spin-phonon BSE becomes
\begin{equation}\label{eq:example_BSE}
\begin{split}
    &\Lambda^z_{11}=1 +\left(\frac{1}{2}-b^2\right)\left(\frac{\Gamma_{1}\Lambda^z_{34}}{-i\Delta+\Gamma_{3}}+\frac{\Gamma_{1}\Lambda^z_{43}}{i\Delta+\Gamma_{3}}\right)\\
    -&\Lambda^z_{22}=-1 -\left(\frac{1}{2}-b^2\right)\left(\frac{\Gamma_{1}\Lambda^z_{34}}{-i\Delta+\Gamma_{3}}+\frac{\Gamma_{1}\Lambda^z_{43}}{i\Delta+\Gamma_{3}}\right)\\
    &\Lambda^z_{34}=1 +\left(\frac{1}{2}-b^2\right)\left(\frac{\Gamma_{3}\Lambda^z_{11}}{\Gamma_{1}}+\frac{\Gamma_{3}\Lambda^z_{22}}{\Gamma_{1}}\right)\\
    &\Lambda^z_{43}=1 +\left(\frac{1}{2}-b^2\right)\left(\frac{\Gamma_{3}\Lambda^z_{11}}{\Gamma_{1}}+\frac{\Gamma_{3}\Lambda^z_{22}}{\Gamma_{1}}\right).
\end{split}
\end{equation}
Note that this equation treats the phonon-induced spin flips and the spin precessional dynamics self-consistently. %
\\
\indent
The first two lines in Eq.~(\ref{eq:example_BSE}) give  $\Lambda^z_{11}=\Lambda^z_{22}$,  
while from the third and fourth lines we obtain $\Lambda^z_{34}=\Lambda^z_{43}$. 
Therefore, the solution of the spin-phonon BSE is
\begin{equation}\label{eq:Lambda}
\begin{split}
&\Lambda^z_{11} =\Lambda^z_{22}= \frac{\Gamma_{3}^2+\Delta^2+\Gamma_{3}\Gamma_{1}(1-2b^2)}{\Gamma_{3}^2 4b^2(1-b^2)+\Delta^2}\\
&\Lambda^z_{34} =\Lambda^z_{43}= \frac{(\Gamma_{3}^2+\Delta^2)(\Gamma_{1}+\Gamma_{3}(1-2b^2))}{(\Gamma_{3}^2 4b^2(1-b^2)+\Delta^2)\Gamma_{1}}.
\end{split}
\end{equation}
The state-dependent spin relaxation times $\tau^z_{11}$, $\tau^z_{22}$, $\tau^z_{34}$, and $\tau^z_{43}$ are obtained using Eq.~(\ref{eq:dressed_time}):
\begin{equation}\label{eq:srt_model}
\begin{split}
    &\tau^z_{11}=\tau^z_{22}=\frac{\Gamma_{3}^2+\Delta^2+\Gamma_{3}\Gamma_{1}(1-2b^2)}{(\Gamma_{3}^2 4b^2(1-b^2)+\Delta^2)\Gamma_{1}}\\
    &\tau^z_{34}=\tau^z_{43}=\frac{(\Gamma_{3}^2+\Delta^2)(\Gamma_{1}+\Gamma_{3}(1-2b^2))}{(\Gamma_{3}^2 4b^2(1-b^2)+\Delta^2)\Gamma_{1}\Gamma_{3}}.
\end{split}
\end{equation}
\\
\indent
Figure~\ref{fig:srt_result}(a) shows the vertex correction for the lowest state, $\Lambda_{11}^z$ in Eq.~(\ref{eq:Lambda}), and Fig.~\ref{fig:srt_result}(b) shows the spin relaxation time ${\tau_{11}^z}$ from Eq.~(\ref{eq:srt_model}), both plotted as a function of the $e$-ph collision time $\tau_1=1/\Gamma_1$. %
These quantities show three distinct regimes with a qualitatively different dependence on the $e$-ph collision time, which correspond to the EY, DP and strong-precession regimes.
Our formalism encompasses these three regimes~\cite{parkPredicting2022} %
because it can capture both spin-flip scattering and spin precession. %
The physics of these regimes is discussed below.
\subsection{Elliott-Yafet regime}
\vspace{-8pt}
We first focus on the EY regime, where spin relaxation occurs primarily through spin-flip transitions. 
In this regime, the spin-flip probability is small since $b\ll1$, and the spin precession rate is much smaller than the scattering rate, $\Delta \ll 2\Gamma_3 b$. 
Using these conditions in Eq.~(\ref{eq:Lambda}), we obtain the following vertex corrections in the EY regime:
\begin{equation}\label{eq:lambda_ey}
\begin{split}
    \Lambda^z_{11}=\Lambda^z_{22}= \frac{\tau_1+\tau_3}{4b^2}\frac{1}{\tau_1}\\
    \Lambda^z_{34}=\Lambda^z_{43}=\frac{\tau_1+\tau_3}{4b^2}\frac{1}{\tau_3},
\end{split}
\end{equation}
where $\tau_{1}=1/\Gamma_{1}$ and $\tau_{3}=1/\Gamma_{3}$ are state-dependent $e$-ph collision times.
These vertex corrections determine the spin relaxation times via Eq.~(\ref{eq:dressed_time_spin}):
\begin{equation}\label{eq:srt_ey}
    \tau^z_{11}=\tau^z_{22}=\tau^z_{34}=\tau^z_{43}=\frac{\tau_{1}+\tau_{3}}{4b^2}.
\end{equation}
These results, shown in Fig.~\ref{fig:srt_result}, approximate well the solution of the spin-phonon BSE in the EY regime. %
\\
\indent
In the conventional theory of EY spin relaxation, the spin relaxation times are proportional to $1/b^2$ and to the $e$-ph collision times (here, $\tau_1$ and $\tau_3$)~\cite{parkSpinphonon2020,baralReexamination2016,vollmarGeneralized2017}.
Our results in Eq.~(\ref{eq:srt_ey}) are consistent with that trend, although the spin relaxation times are proportional to the average of the $e$-ph collision times of the two levels, $(\tau_1 + \tau_3)/2$, and not to their individual values. 
This difference is a result of considering both forward- and back-scattering processes between the two electronic levels in the spin-phonon BSE~\cite{kimVertex2019}, different from the simpler sRTA~\cite{parkSpinphonon2020,baralReexamination2016,vollmarGeneralized2017} which neglects electron back-scattering.\\
\subsection{Dyakonov-Perel regime}
\vspace{-8pt}
Next, we discuss the regime where spin precession is important and governs spin relaxation.
Here, the spin-flip probability is still small ($b\ll1$) but the internal magnetic field due to SOC is significant, such that $\Delta \gg 2\Gamma_3 b$.
Inserting these conditions in Eqs.~(\ref{eq:Lambda})-(\ref{eq:srt_model}), the vertex corrections become
\begin{equation}\label{eq:lambda_precession}
\begin{split}
    &\Lambda^z_{11}=\Lambda^z_{22}=\frac{\Gamma_{3}^2+\Gamma_{3}\Gamma_{1}+\Delta^2}{\Delta^2},\\
    &\Lambda^z_{34}=\Lambda^z_{43}=\frac{(\Gamma_{3}^2+\Delta^2)(\Gamma_1+\Gamma_3)}{\Delta^2\Gamma_1},
\end{split}
\end{equation}
and for the spin relaxation times we obtain
\begin{equation}\label{eq:srt_precession}
\begin{split}
    &\tau^z_{11}=\tau^z_{22}=\frac{\Gamma_{3}^2+\Gamma_{3}\Gamma_{1}+\Delta^2}{\Delta^2\Gamma_1},\\
    &\tau^z_{34}=\tau^z_{43}=\frac{(\Gamma_{3}^2+\Delta^2)(\Gamma_1+\Gamma_3)}{\Delta^2\Gamma_1\Gamma_3}.
\end{split}
\end{equation}
These results describe spin relaxation governed by spin precession and renormalized by the $e$-ph interactions.
\\
\indent 
In the DP regime, the $e$-ph scattering rates are much greater than the bare spin precession rate $\Delta$, and thus $\Gamma_{1,3} \gg  \Delta$. The vertex corrections in the DP regime become
\begin{equation}\label{eq:lambda_dp}
\begin{split}
    &\Lambda^z_{11}=\Lambda^z_{22}=\frac{\Gamma_{3}^2+\Gamma_{3}\Gamma_{1}}{\Delta^2\Gamma_1}\Gamma_1\\
    &\Lambda^z_{34}=\Lambda^z_{43}=\frac{\Gamma_{3}^2+\Gamma_{3}\Gamma_{1}}{\Delta^2\Gamma_1}\Gamma_3,
\end{split}
\end{equation}
while for the spin relaxation times we obtain
\begin{equation}\label{eq:srt_dp}
    \tau^z_{11}=\tau^z_{22}=\tau^z_{34}=\tau^z_{43}=\frac{\Gamma_{3}^2+\Gamma_{3}\Gamma_{1}}{\Delta^2\Gamma_1}.
\end{equation}
These results, shown in Fig.~\ref{fig:srt_result}, are an excellent approximation to the BSE solution in the DP regime.
\\
\indent
A hallmark of DP relaxation is the inverse proportionality between the spin relaxation and $e$-ph collision times~\cite{zuticSpintronics2004}, a trend captured by our treatment of the DP regime [see Eq.~(\ref{eq:srt_dp})]. %
Our formalism shows in Eq.~(\ref{eq:lambda_dp}) that this trend originates from the inverse-square scaling of the vertex corrections with the $e$-ph collision times, $\Lambda \sim \Gamma_{1,3}^2 \sim 1/\tau_{1,3}^2 $. %
Note that in the EY regime, rescaling the $e$-ph collision times by a constant factor has no effect on the vertex corrections, which depend only on the ratio $\tau_1/\tau_3$, and thus the EY spin relaxation times are proportional to the $e$-ph collision times. The situation is different in the DP regime, where rescaling the $e$-ph collision times changes the vertex corrections. 
These scaling trends can be employed in real materials to identify the dominant microscopic mechanisms for spin relaxation and decoherence~\cite{parkPredicting2022}.
\\
\indent
\subsection{Strong-precession regime}
\vspace{-8pt}
A third, distinct regime is realized when the Zeeman splitting is much greater than the $e$-ph scattering rates, $\Delta \gg  \Gamma_{1,3} $, such that the spins precess much faster than the rate of $e$-ph collisions.
In this strong-precession regime~\cite{burkovSpin2004,szolnokiSpinrelaxation2017,csoszGeneric2020}, the vertex corrections become
\begin{equation}\label{eq:lambda_strong}
\begin{split}
    &\Lambda^z_{11}=\Lambda^z_{22}=1,\\
    &\Lambda^z_{34}=\Lambda^z_{43}=1+\frac{\Gamma_{3}}{\Gamma_1}
\end{split}
\end{equation}
and can be approximated as  $\Lambda\approx1$.
Therefore, the vertex corrections are less important than in the EY or DP regimes.
It follows that in the strong-precession regime the spin relaxation times are proportional to the $e$-ph collision times:
\begin{equation}\label{eq:srt_opposite_dp}
\begin{split}
    &\tau^z_{11}=\tau^z_{22}=\frac{1}{\Gamma_1},\\
    &\tau^z_{34}=\tau^z_{43}=\frac{1}{\Gamma_1}+\frac{1}{\Gamma_3}.
\end{split}
\end{equation}
These results, shown in Fig.~\ref{fig:srt_result}, approximate well the BSE solution in the strong-precession regime. 
Their physical interpretation is interesting: in the strong-precession regime, the spins precess for many full cycles between phonon collisions, randomizing the spin direction. Consequently, the spin relaxation times become equal to the $e$-ph collision times, and thus the vertex corrections $\Lambda\approx 1$ can be neglected.
\newpage
\subsection{Uncovering the three regimes in GaAs}
\vspace{-8pt}
To identify the three spin relaxation mechanisms in a real material, we study spin relaxation in GaAs using our first-principles implementation of the spin-phonon BSE~\cite{parkPredicting2022}. 
We focus on spin relaxation for conduction band electrons in GaAs at room temperature (300~K), and investigate how the spin relaxation times depend on the $e$-ph collision times. 
\\
\indent
We compute the ground state and band structures of GaAs with density functional theory (DFT), using a plane-wave basis in the {\sc Quantum ESPRESSO} code~\cite{giannozziQUANTUM2009} 
and employing the HSE06 hybrid functional~\cite{heydHybrid2003} to obtain an accurate band gap and electronic structure.
We use fully relativistic norm-conserving pseudopotentials, generated in the local-density approximation (LDA) with {\sc Pseudo Dojo}~\cite{vansettenPseudoDojo2018}, together with a kinetic energy cutoff of 72~Ry and a relaxed lattice constant of 5.60~$\text{\AA}$.
The phonon energies and perturbation potentials are computed using density functional perturbation theory (DFPT)~\cite{baroniPhonons2001}. 
Using our {\sc perturbo} code~\cite{zhouPerturbo2021}, we compute the $e$-ph matrix elements on coarse Brillouin zone grids with $8\times8\times8$ $\kk$ and $\qq$ points,
following which we generate spinor Wannier functions with the {\sc Wannier90} code~\cite{pizziWannier902020} and use them in {\sc perturbo}~\cite{zhouPerturbo2021}, with a method we developed in Ref.~\cite{parkSpinphonon2020}, to jointly interpolate the $e$-ph matrix elements and spin matrices. 
The long-range quadrupole $e$-ph interactions~\cite{jhalaniPiezoelectric2020,parkLongrange2020,bruninElectronphonon2020,bruninPhononlimited2020} are included to fully account for $e$-ph coupling with long-wavelength phonons. We interpolate the $e$-ph matrix elements to fine Brillouin zone grids with up to $200\times200\times200$ $\kk$ and $\qq$ points, and a 10~meV Gaussian broadening for the energy-conserving delta functions in the $e$-ph scattering rates~\cite{zhouInitio2016}. 
The spin relaxation times in Eq.~(\ref{eq:srt_def}) are computed using the tetrahedron integration method~\cite{blochlImproved1994}, assuming a non-degenerate electron concentration of $10^{16}~\text{cm}^{-3}$.  
The spin-phonon BSE in Eq.~(\ref{eq:bse_spin}) %
is solved with an augmented iterative approach described in Ref.~\cite{parkPredicting2022}.
\\
\indent
\begin{figure}[!t]
\includegraphics[width=0.9\columnwidth]{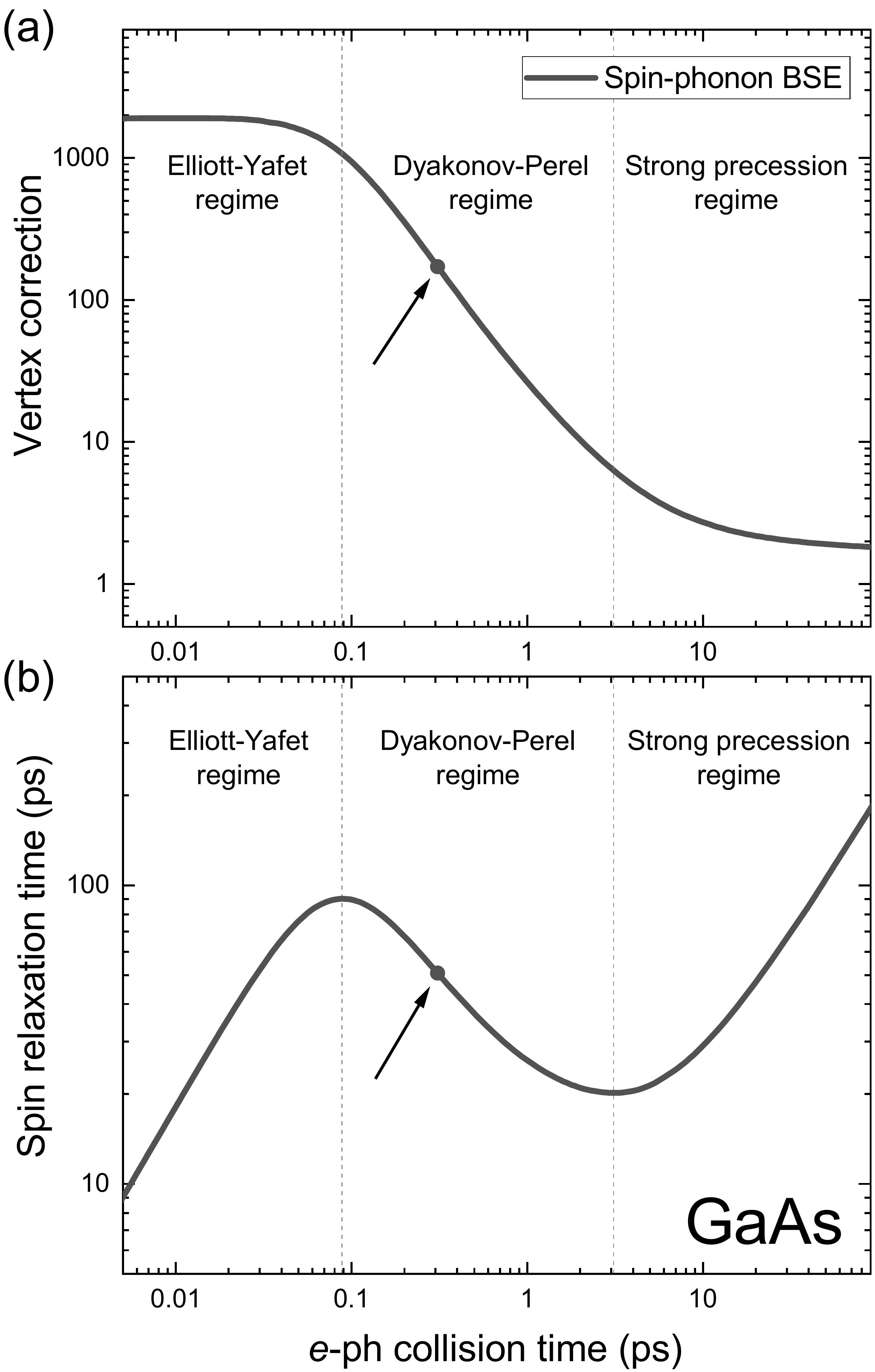}
\caption{First-principles calculations of (a) vertex corrections and (b) spin relaxation times for electron spins in GaAs, computed at room temperature and plotted as a function of the average $e$-ph collision time. 
The arrows indicate the values for the real material, GaAs (black dot), while the other values are obtained by rescaling the $e$-ph interaction strength, as explained in the text. 
The vertical dotted lines are placed at the inflection points of the spin relaxation times, and separate the EY, DP, and strong-precession regimes.
}\label{fig:gaas}
\end{figure}
Using our first-principles spin-phonon BSE, we compute the spin relaxation time for electron spins in GaAs, and obtain a value 51~ps at 300~K in excellent agreement with the experimental value of 42~ps~\cite{oertelHigh2008}. %
We also obtain an average $e$-ph collision time of 410 fs, computing using 
\begin{equation}
\langle\tau^\text{e-ph}\rangle = \frac{\sum_{n\kk} \tau^{\text{e-ph}}_{nn\kk}(\varepsilon_{n\kk}) (-\frac{df_{n\kk}}{d\varepsilon})}{\sum_{n\kk}   
    \,\big(\!-\frac{df_{n\kk}}{d\varepsilon} \big)}
\end{equation}
 and an average vertex correction $\langle \Lambda^z \rangle \!\approx\! 171$, computed from
\begin{equation}\label{eq:vc_avg}
    \langle \Lambda^z \rangle=\frac{\sum_{n\kk} \abs{s^z_{nn\kk}}^2\Lambda^{z}_{nn\kk}(\varepsilon_{n\kk}) (-\frac{df_{n\kk}}{d\varepsilon})}{\sum_{n\kk}   \abs{s^z_{nn\kk}}^2
    \,\big(\!-\frac{df_{n\kk}}{d\varepsilon} \big)}.
\end{equation}
These ``real-material'' values for GaAs are shown with a dot in Fig.~\ref{fig:gaas}(a) and (b), where we also show spin relaxation times and vertex corrections obtained by artificially varying the average $e$-ph collision time (by rescaling the $e$-ph matrix elements). The resulting trends show that the spin relaxation times are inversely proportional to the $e$-ph collision time, placing GaAs in the DP spin-relaxation regime. 
\\
\indent
Upon decreasing the average $e$-ph collision time $\langle \tau^\text{e-ph} \rangle$ in Fig.~\ref{fig:gaas}(a), the system evolves from the DP to the EY regime, and the vertex correction first increases and then saturates to a maximal value of $\sim$1900 in the short $e$-ph collision time limit. 
As a consequence, the spin relaxation time in Fig.~\ref{fig:gaas}(b) peaks at $\langle \tau^\text{e-ph}\rangle \!\approx\! 0.09$~ps and then decreases linearly at shorter $e$-ph collision times. 
This crossover from the DP to the EY mechanism, obtained here by artificially tuning the $e$-ph collision time in GaAs, is consistent with the results from our model system. 
\\
\indent
Conversely, when the $e$-ph collision time is increased, the system transitions from the DP to the strong-precession regime: 
Fig.~\ref{fig:gaas}(a) shows that the vertex correction first decreases as $\langle\tau^{\text{e-ph}}\rangle^{-2}$ and then plateaus to a minimal value close to unity for long $e$-ph collision times.
Accordingly, the spin relaxation time in Fig.~\ref{fig:gaas}(b) reaches a minimum for $\langle\tau^\text{e-ph}\rangle\approx3$~ps and then increases linearly at longer $e$-ph collision times. 
\\
\indent
In summary, real materials exhibit the same spin relaxation mechanisms and vertex correction trends as our two-level system. 
These regimes for phonon-induced spin dynamics are summarized in Table~\ref{table:eydp}.
The presence of three distinct spin-relaxation regimes is general, and we expect it to be valid beyond the case of a simple semiconductor (GaAs) studied here.
\begingroup
\setlength{\tabcolsep}{4pt} %
\renewcommand{\arraystretch}{1.5} %
\begin{table}[!t]
\centering
\caption{
Summary of the Elliott-Yafet, Dyakonov-Perel, and strong-precession regimes.
}
\begin{tabular}{c c c c} 
 \hline\hline
  ~ & EY & DP & Strong-precession \\ [0.5ex] 
 \hline
   & Spin-flip               & Spin precession & Spin precession    \\
  $\Lambda$ & $ \sim{1}/{b^2}$   & $ \sim{1}/{(\tau^{\text{e-ph}})^2}$ & $ \sim1$ \\ %
  $\tau^s$ & $ \sim{\tau^{\text{e-ph}}}/{b^2}$     & $ \sim{1}/{\tau^{\text{e-ph}}}$ & $ \sim\tau^{\text{e-ph}}$   \\
  [0.5ex]
 \hline\hline
\end{tabular}
\label{table:eydp}
\end{table}
\endgroup
\section{Discussion}
\vspace{-8pt}
Our formalism can capture all three of the EY, DP and strong-precession regimes in a unified framework. 
The reason can be inferred from the diagrammatic representation of the spin-phonon BSE in Fig.~\ref{fig:bubble_bse}(c). 
In the diagrams, the EY relaxation is due to $e$-ph scattering in the presence of spin mixing, which originates from the $e$-ph interactions $\left[g_{n'm'\nu}(\kk,\qq)\right]^*$ and $g_{nm\nu}(\kk,\qq)$ and the wiggly line in the kernel of the BSE. 
The DP and strong-precession mechanisms are included by virtue of the electron propagators with two different band indices, $\mathcal{G}_{ml\kk+\qq}\mathcal{G}_{l'm'\kk+\qq}$, placed between the wiggly line and the spin vertex. These propagators take into account spin precession (due to the SOC field) between $e$-ph collisions. This elegant formalism captures a  wide range of spin physics in a single diagram.
\\
\indent
Although our discussion has focused on $T_1$ spin relaxation times, the spin decoherence times $T_2$ at finite magnetic fields can also be computed, as we plan to show in future work. Finally, the approach presented in this work for the phonon-dressed vertex is general and goes beyond spin relaxation. It can be employed to study the dynamics of any observable that couples with phonons, for which we also expect to find the three phonon-induced relaxation regimes discussed above for spin dynamics. 
\section{Conclusion}
\vspace{-8pt}
We have formulated a theory for the vertex corrections from $e$-ph interactions to the susceptibility of a non-diagonal operator.  
The key result is a self-consistent BSE to calculate the phonon-dressed vertex, which encodes the dynamics of the operator coupling with phonons. When applied to spin, this approach enables quantitative calculations of spin relaxation and decoherence~\cite{parkPredicting2022}. 
We have shown that our spin-phonon BSE captures both spin-flip transitions and spin precession, unifying the treatment of three spin decoherence mechanisms (EY, DP, and strong-precession) conventionally treated with separate heuristic models. 
By leveraging efficient workflows for first-principles $e$-ph calculations~\cite{zhouPerturbo2021}, 
our method enables quantitative studies of spin relaxation and decoherence in a wide range of bulk and two-dimensional materials, as we show in the companion paper~\cite{parkPredicting2022}. 
These advances open new avenues for understanding spin relaxation and decoherence in spintronics, magnetism, multiferroics, quantum materials and quantum technologies.
\\
\begin{acknowledgments}
\vspace{-10pt}
This work was supported by the National Science Foundation under Grants No. DMR-1750613 and QII-TAQS 1936350, which provided for method development, and Grant No. OAC-2209262, which provided for code development. J.P. acknowledges support by the Korea Foundation for Advanced Studies.
\end{acknowledgments}
\appendix
\section{Ward identity}\label{sec:ward}
\vspace{-8pt}
We derive a Ward identity for our BSE in Eq.~(\ref{eq:bse_full_G}). The Ward identity relates the vertex corrections with the electron self-energy~\cite{kimVertex2019,mahanManyParticle2000,wardIdentity1950}, and guarantees that diagrams are taken into account consistently in the self-energy and in the BSE for the  vertex~\cite{kimVertex2019,mahanManyParticle2000,wardIdentity1950}. 
\\
\indent
For a system with Hamiltonian $H$, the operator $\hat{A}$ is related to the the negative derivative of the Hamiltonian with respect to the external field $\mathcal{F}$,
    $\hat{A} = -\grad_{\mathcal{F}}H$.
We compute the change of the Fan-Migdal self-energy in Eq.~(\ref{eq:sigma}) with respect to $\mathcal{F}$, by taking the derivative
\vspace{-5pt}
\begin{equation}\label{eq:dSE_dG}
\begin{split}
&{\grad_{\mathcal{F}} \Sigma_{nn'\kk}(i\omega_a) } \\
&~~= -\frac{1}{\beta N_q V_{\text{uc}}}\sum_{mm'll'\qq\nu,iq_c} \left[g_{n'm'\nu}(\kk,\qq)\right]^* g_{nm\nu}(\kk,\qq) \\
&~~~ \times\mathcal{D}_{\nu\qq}(iq_c)  \left(  \grad_{\mathcal{F}} \mathcal{G}_{mm'\kk}(i\omega_a+iq_c) \right).
\end{split}
\end{equation}
Employing the matrix identity
\begin{equation}
\begin{split}
    {\grad_{\mathcal{F}} \mathcal{G}_{mm'\kk}} &=- \sum_{ll'} \mathcal{G}_{ml\kk}
    \left[ {\grad_{\mathcal{F}} (\mathcal{G}^{-1}}) \right]_{ll'\kk}\mathcal{G}_{l'm'\kk} \\
    &= -\sum_{ll'} \mathcal{G}_{ml\kk}
    \left(  - \grad_{\mathcal{F}}H_{ll'\kk} - \grad_{\mathcal{F}} \Sigma_{ll'\kk}  \right)\mathcal{G}_{l'm'\kk},
\end{split}
\end{equation}
we obtain a self-consistent equation for the self-energy derivatives:
\begin{equation}\label{eq:dSE}
\begin{split}
&{\grad_{\mathcal{F}} \Sigma_{nn'\kk}(i\omega_a) } \\
&~~= -\frac{1}{\beta N_q V_{\text{uc}}}\sum_{mm'll'\qq\nu,iq_c} \left[g_{n'm'\nu}(\kk,\qq)\right]^* g_{nm\nu}(\kk,\qq) \\
&~~~ \times\mathcal{D}_{\nu\qq}(iq_c) \mathcal{G}_{ml\kk+\qq}(i\omega_a+iq_c) \mathcal{G}_{l'm'\kk+\qq}(i\omega_a+iq_c) \\
&~~~ \times \left(  \grad_{\mathcal{F}}H_{ll'\kk} + \grad_{\mathcal{F}} \Sigma_{ll'\kk}(i\omega_a+iq_c)  \right).
\end{split}
\end{equation}
\\
\indent
Comparing Eq.~(\ref{eq:dSE}) and Eq.~(\ref{eq:bse_qc}) in the $i\nu_b \to 0$ limit, we discover the Ward identity expressed in terms of the Hamiltonian, vertex corrections, and self-energy:
\begin{equation}\label{eq:Ward}
(\grad_{\mathcal{F}}H)\bm{\Lambda}(i\omega_a,i\omega_a) = \grad_{\mathcal{F}}H + {\grad_{\mathcal{F}} \Sigma(i\omega_a) },
\end{equation}
where $(\grad_{\mathcal{F}}H)\bm{\Lambda}(i\omega_a,i\omega_a)=(\frac{\partial H_{nn'\kk}}{{\mathcal{\partial F^\alpha}}}){\Lambda}^\alpha_{nn'\kk}(i\omega_a,i\omega_a)$.
Equation~(\ref{eq:Ward}) can be equivalently expressed in terms of the operator matrix elements $A^\alpha_{nn'\kk}$\,, 
\begin{equation}\label{eq:Ward_A}
    A^\alpha_{nn'\kk}\Lambda^\alpha_{nn'\kk}(i\omega_a,i\omega_a) = A^\alpha_{nn'\kk} - \frac{\partial \Sigma_{nn'\kk}(i\omega_a) } {\partial \mathcal{F}^\alpha}.
\end{equation}
\\
\indent
Our expression for the Ward identity in Eqs.~(\ref{eq:Ward})-(\ref{eq:Ward_A}) is consistent with the Ward identity for the velocity operator derived in Ref.~\cite{kimVertex2019}:
\begin{equation}\label{eq:Ward_velocity}
v^\alpha_{nn\kk}\Lambda^\alpha_{nn\kk}(i\omega_a,i\omega_a) = v^\alpha_{nn\kk} + \frac{\partial \Sigma_{nn\kk}(i\omega_a) } {\partial k^\alpha},
\end{equation}
as the velocity operator is defined as $\hat{v} = \grad_{\mathcal{F}} H $ with $\mathcal{F}=\kk$.
This result further validates our BSE for the phonon-dressed vertex.\\ 

\vspace{10pt}
\bibliographystyle{apsrev4-2}

\end{document}